\documentclass[fleqn,usenatbib]{mnras}

\usepackage{newtxtext,newtxmath}

\usepackage[T1]{fontenc}
\usepackage{ae,aecompl}

\usepackage{graphicx}	
\usepackage{amsmath}	

\usepackage{amssymb}	

\usepackage[table]{xcolor}

\usepackage{lastpage}
\usepackage{natbib}
\usepackage{setspace}
\usepackage[utf8]{inputenc}

\title[Systematics from initial helium abundance]{Asteroseismic modelling of solar-type stars: A deeper look at the treatment of initial helium abundance }

\author[Nsamba et al.]{
Benard Nsamba$^{1,2}$\thanks{E-mail: nsamba@mpa-garching.mpg.de},
Nuno Moedas$^{2,3}$, Tiago L. Campante$^{2,3}$, Margarida S. Cunha$^{2,8}$,
\newauthor
Antonio Garc\'{i}a Hern\'{a}ndez$^{4,7}$,
Juan C. Su\'{a}rez$^{4,7}$, 
M\'{a}rio J. P. F. G. Monteiro$^{2,3}$, 
\newauthor
Jo\~{a}o Fernandes$^{5}$, Chen Jiang$^{6}$ and Babatunde Akinsanmi$^{2,3}$
\\
$^{1}$ Max-Planck-Institut f\"{u}r Astrophysik, Karl-Schwarzschild-Str. 1, D-85748 Garching, Germany\\
$^{2}$Instituto de Astrof\'{\i}sica e Ci\^{e}ncias do Espa\c{c}o, Universidade do Porto,  Rua das Estrelas, PT4150-762 Porto, Portugal\\
$^{3}$Departamento de F\'{\i}sica e Astronomia, Faculdade de Ci\^{e}ncias da Universidade do Porto, Rua do Campo Alegre, s/n, PT4169-007\\ Porto, Portugal\\
$^{4}$Department of Theoretical Physics and Cosmology, University of Granada (UGR), E-18071Granada, Spain\\
$^{5}$Univ Coimbra, CITEUC, Department of Mathematics, Geophysical and Astronomical Observatory, Portugal\\
$^{6}$School of Physics and Astronomy, Sun Yat-sen University, 2 Daxue Road, Tangjia, Zhuhai 519082, Guangdong Province, China\\
$^{7}$Instituto de Astrof\'isica de Andaluc\'ia (CSIC), Glorieta de la Astronom\'ia S/N, 18008, Granada, Spain\\
$^{8}$School of Physics and Astronomy, University of Birmingham, Birmingham, B15 2TT, United Kingdom\\
}

\date{Accepted 2020 October 12. Received 2020 October 8; in original form 2020 August 10}

\pubyear{2020}

\begin{document}
\label{firstpage}
\pagerange{\pageref{firstpage}--\pageref{lastpage}}
\maketitle

\begin{abstract}
Detailed understanding of stellar physics is essential towards a robust determination of stellar properties (e.g. radius, mass, and age). Among the vital input physics used in the modelling of solar-type stars which remain poorly constrained, is the initial helium abundance. To this end, when constructing stellar model grids,  the initial helium abundance is estimated either (i) by using the semi-empirical helium-to-heavy element enrichment ratio, (${\Delta Y}/{\Delta Z}$), anchored to the standard Big Bang Nucleosynthesis value or (ii) by setting the initial helium abundance as a free variable.
Adopting 35 low-mass, solar-type stars with multi-year {\it Kepler} photometry from the asteroseismic ``LEGACY'' sample, we explore the systematic uncertainties on the inferred stellar parameters (i.e., radius, mass, and age) arising from the treatment of the initial helium abundance in stellar model grids .
The stellar masses and radii derived from grids with free initial helium abundance are lower compared to those from grids based on a fixed ${\Delta Y}/{\Delta Z}$ ratio. We find the systematic uncertainties on mean density, radius, mass, and age arising from grids which employ a fixed value of ${\Delta Y}/{\Delta Z}$ and those with free initial helium abundance to be $\sim$ 0.9\%, $\sim$ 2\%, $\sim$ 5\% and $\sim$ 29\%, respectively. We report that the systematic uncertainties on the inferred masses and radii arising from the treatment of initial helium abundance in stellar grids lie within the expected accuracy limits of ESA's {\it{PLATO}}, although this is not the case for the age.

 \end{abstract}



\section{Introduction}
\label{intro}
One of the vital model inputs employed in stellar evolution codes (such as, MESA\footnote{Modules for Experiments in Stellar Astrophysics}; \citealt{Pax2,Pax3,Pax4}, GARSTEC\footnote{Garching Stellar Evolution Code}; \citealt{Weiss2008}, YREC\footnote{Yale Rotating Stellar Evolution Code}; \citealt{2008ADemarque}) aiding our understanding of the formation, structure, and evolution of stars are their chemical compositions. When modelling an ensemble of solar-type stars, solar metallicity mixtures are commonly adopted. This is because a good agreement exists between element abundances of the Sun and solar-type stars. However, at low metallicities, the solar-type stars get more enhanced in some elements (alpha elements) when compared to the Sun (see \citealt{2012Adib} for details). Based on the approach used in determining solar abundances (e.g. through the analysis of solar photospheric spectrum and meteorite) significant discrepancies exist among different surface elements \citep{Anders1989,Grevesse1998,Lodders2003,Grevesse2007,Asplund2009}. Using spectroscopic methods, estimation of helium element abundance in solar-type stars still present vital challenges. This is because the envelope temperatures of these stars are not sufficient to excite helium, thus few or no helium lines are detectable in their spectra using spectroscopic observations. This is the origin of the challenges faced by stellar modellers in determining initial helium abundances to be used in stellar evolution models.

In order to circumvent the initial helium abundance problem when constructing stellar models, there are two approaches commonly adopted:
(i) determining the initial helium abundance ($Y_i$) using the helium-to-heavy element enrichment ratio ($\Delta Y / \Delta Z$) anchored to the primordial Big Bang nucleosynthesis values (i.e.,  $Z_{0}$ = 0 and $Y_{0}$ = $0.2484$: \citealt{Cyburt}) through the following expression;
\begin{equation}
 \left(\frac{\Delta Y }{\Delta Z} \right) = \frac{Y_i - Y_{0}}{Z_i - Z_0}~~, 
 \label{helium}
\end{equation}
where $Z_i$ is the initial metal abundance. This has been widely adopted in stellar grid construction, e.g. \citet{Basu_2010,Metcalfe_2010,Gai_2011,Lebreton2014,Metcalfe2014,Aguirre,Aguirre1,Rodrigues,2018Frandsen,Nsambab,Nsambaa,2019Nsamb,2019Verma,Li,2020arSerenelli,2020Jiang} among others.
(ii) Setting free values of $Y_i$. These may be set to range between 0.22 to 0.34 (e.g. \citealt{Mathur_2012,Metcalfe2014,Verma_2014,Aguirre, Aguirre1,2019Bellinger,2020Angelou}) or a low bound on $Y_i$ may be set corresponding to the primordial Big Bang nucleosynthesis value (e.g. \citealt{Aguirre,Aguirre1,Li,2020Valle}).
The former is known to yield optimal model solutions inferred using forward modelling techniques\footnote{Forward modeling involves determining fundamental stellar parameters by matching model parameters to the observed properties, e.g. seismic observables (such as individual oscillation frequencies or frequency ratios) and non seismic observables (effective temperature, metallicity, luminosity)} having initial helium values that in some cases are below the primordial Big Bang nucleosynthesis value. (e.g. \citealt{Bonaca2012,Lebreton2014,Metcalfe2014,Aguirre,Aguirre1}).
This option of determining $Y_i$ leads to an increase in the number of stellar grid variable parameters, thus it is costly in terms of the computational time and storage capacity for the computed models. 

The helium enrichment law in Eq.~(\ref{helium}) is commonly employed in stellar model grid construction, however, a major setback in using Eq.~(\ref{helium}) is that no consensus has yet been reached regarding the value of the helium-to-heavy element ratio.
Based on the observation of K dwarf stars in the Hipparcos catalog, \citet{Jimenez} reported  the $\Delta Y / \Delta Z$ value to be $2.1 \pm 0.4 $.  These results are in agreement with those obtained based on a set of Padova isochrones constructed with a wide range of helium and metal content to fit observations of nearby K dwarf stars, i.e., $\Delta Y / \Delta Z = 2.1 \pm 0.9$ \citep{2007MCasagrand}.
Exploring the metal-poor galaxy H\,II regions, Magellanic cloud H\,II regions, and M17 abundances while taking into account the effects of temperature fluctuations, \citet{Balser} reported the value of $\Delta Y / \Delta Z$ to be $1.6$. Interestingly, when using only galaxy H\,II region S206 and M17, \citet{Balser} determines $\Delta Y / \Delta Z = 1.41 \pm 0.62$, a value reported to be consistent with that from standard chemical evolution models. \citet{Serenelli} reported the helium-to-heavy element enrichment ratio of the Sun to vary in the range $1.7\leq \Delta Y / \Delta Z \leq 2.2$ depending on the choice of solar composition. 
Through the analysis of glitch signatures caused by the ionisation of helium in 38 {\em Kepler} "LEGACY" sample stars, \citet{2019Verma} estimated their surface helium abundances. Combining these values with abundance differences caused by gravitational settling in stellar models, they estimated the initial helium abundances and derived a primordial helium abundance of 0.244\,$\pm$\,0.019 with $\Delta Y / \Delta Z$ = 1.226 $\pm$ 0.843.
In sum, the acceptable values of the helium-to-heavy enrichment ratio span the interval $1 \leq \Delta Y / \Delta Z \leq 3$, notwithstanding lower values being found in some cases, especially when one determines the value of $\Delta Y / \Delta Z$ based on solar calibrations. Furthermore, through exploring the relations between $Y$ and $Z$ in the interstellar medium of simulated disc galaxies, \citet{Vincenzo} reports that $\Delta Y / \Delta Z$ is not constant and evolves as a function of time, depending on the specific chemical element chosen  to trace $Z$. 

Given the fact that the helium abundance affects the structure and evolution life time of stars, this implies that any uncertainties in helium abundance directly impacts on the determination of the stellar parameters such as, radius, mass, and age.  Therefore, the treatment of the initial helium abundance in stellar models is a substantial source of systematic uncertainties on stellar properties derived using forward modelling techniques. 
\citet{Lebreton2014} carried out a detailed characterisation of the CoRoT exoplanet host HD 52265 and reported a scatter of about 5 per cent in mass arising from the treatment of  initial helium mass fraction. Using synthetic data for about 10,000 artificial stars, \citet{Valle2014} found the systematic bias on mass and radius estimation arising from a variation of $\pm 1$ in $\Delta Y / \Delta Z$ to be 2.3 per cent and 1.1 per cent, respectively. Further, \citet{Valle2015} reported the systematic bias in age to  be about one-fourth of the statistical error in the first 30 per cent of the evolution, while its negligible for more evolved stars. 

In this article, we perform a detailed study of a sample of  
low-mass main-sequence stars with high signal-to-noise asteroseismic data from multi-year {\it {Kepler}} photometry \citep{Lund}, making asteroseismic inferences to quantify the systematic uncertainties on the derived stellar global parameters (i.e., mean density, radius, mass, and age) arising from the treatment of initial helium abundance in stellar grids. In addition, we assess if these systematic uncertainties are within the ESA's PLATO (PLAnetary Transits and Oscillations of stars; \citealt{2014Benz}) mission accuracy requirements for exoplanet-host star properties, needed for precise planet characterisation.

This article is organised as follows.  In Sect.~\ref{sample}, we describe our target sample, the details of the stellar grids, and optimisation routines employed.
In Sect.~\ref{results}, we present our results and discussions while Sect.~\ref{con} contains our conclusions.

\section{Target Sample and stellar models}
\label{sample}
The seismic and atmospheric observations used for our target sample are described in Sect.~\ref{target} while Sect.~\ref{models} is composed of a description of the stellar grids and best-fit model selection procedures employed in the forward modelling routines.
\subsection{{\em Kepler} data}
\label{target}
\begin{figure}
        \includegraphics[width=\columnwidth]{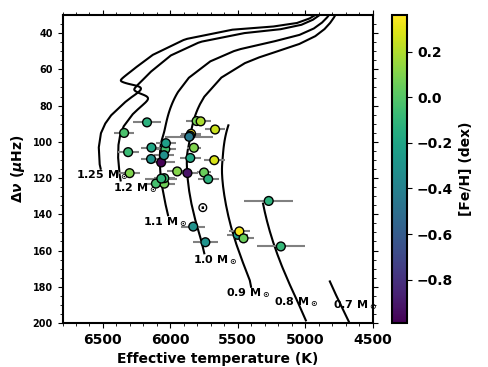}
      \caption{Asteroseismic Hertzsprung-Russell diagram showing the parameter
space covered by 35 low-mass LEGACY sample stars. Circles: target stars colour coded according to their respective metallicities. Black lines: stellar evolutionary tracks ranging in mass from 0.7 $\rm M_\odot$ to 1.25 $\rm M_\odot$ constructed using a mixing length parameter ($\alpha_{\rm mlt}$) of 1.8 in order to describe convection, a heavy metal mass fraction ($Z$) of 0.02, and an initial helium mass fraction was obtained using Eq. (\ref{helium}), with $\Delta Y / \Delta Z$ set to 1.4.
      }
    \label{tracks}
\end{figure}

The location of the {\it Kepler} LEGACY sample stars employed in this study is shown in the asteroseismic Hertzsprung-Russel diagram (see Fig.~\ref{tracks}). 
The y-axis of Fig.~\ref{tracks} contains the values of the average large separations, $\Delta \nu$, adopted from \citet{Lund}. These were estimated in that work using as a  linear fit to the spherical mode degree $l = 0$ frequencies expressed as a function of the radial order, $n$. 

The sample shown in Fig.~\ref{tracks} consists of 35 low-mass solar-type stars  with at least 12 months observation of short cadence data ($\Delta t$ = 58.89 s). The seismic data of these stars are available in \citet{Lund}.  It is worth noting that for each star, we only adopted oscillation modes whose probability of detection is reported to be at least ``{\it{strong}}'' by \citet{Lund}. Refer to Equation~15 in \citet{Lund} for details on the probability of the detection of the given set of modes.
\begin{table}
 
\caption{Global asteroseismic and spectroscopic parameters of our stellar sample.}
\label{1}
\begin{tabular}{ ccccc} 
\hline \hline 
KIC  &   $\Delta \nu$ ($\mu$ Hz)&	$T_{\mbox{eff}}$ (K)	&  [Fe/H] (dex)	& Ref.\\
\hline 
3427720	& 120.068$^{+0.031}_{-0.032}$	&   6045 $\pm$ 77		&  -0.06 $\pm$ 0.10 &   1  	\\
3656476	& 93.194$^{+0.018}_{-0.020}$	&	5668 $\pm$ 77		&   0.25 $\pm$ 0.10 &   1	\\
3735871	& 123.049$^{+0.047}_{-0.046}$	&	6107 $\pm$ 77		&  -0.04 $\pm$ 0.10 &   1	 \\
4914923	& 88.531$^{+0.019}_{-0.019}$	&	5805 $\pm$ 77		&   0.08 $\pm$ 0.10 &   1	\\
5184732	& 95.545$^{+0.024}_{-0.023}$	&	5846 $\pm$ 77		&   0.36 $\pm$ 0.10 &   1	\\
5950854	& 96.629$^{+0.102}_{-0.107}$	&   5853 $\pm$ 77		&  -0.23 $\pm$ 0.10 &   1	\\
6106415	& 104.074$^{+0.023}_{-0.026}$	&	6037 $\pm$ 77		&  -0.04 $\pm$ 0.10 &   1	\\
6116048	& 100.754$^{+0.017}_{-0.017}$	&	6033 $\pm$ 77		&  -0.23 $\pm$ 0.10 &   1	\\
6225718	& 105.695$^{+0.018}_{-0.017}$	&	6313 $\pm$ 77		&  -0.07 $\pm$ 0.10 &   1	\\
6603624	& 110.128$^{+0.012}_{-0.012}$	&	5674 $\pm$ 77		&  0.28  $\pm$ 0.10 &   1	\\
7106245	& 111.376$^{+0.063}_{-0.061}$	&	6068 $\pm$ 102  	&  -0.99 $\pm$ 0.19 &   2	\\
7296438	& 88.698$^{+0.040}_{-0.036}$	&	5775 $\pm$ 77		&  0.19  $\pm$ 0.10 &   1	\\
7871531	& 151.329$^{+0.025}_{-0.023}$	&	5501 $\pm$ 77		&  -0.26 $\pm$ 0.10 &   1	\\
8006161	& 149.427$^{+0.015}_{-0.014}$	&	5488 $\pm$ 77		&  0.34  $\pm$ 0.10 &   1	\\
8150065	& 89.264$^{+0.134}_{-0.121}$	&	6173 $\pm$ 101  	&  -0.13 $\pm$ 0.15 &   2   \\
8179536	& 95.090$^{+0.058}_{-0.054}$	&	6343 $\pm$ 77		&  -0.03 $\pm$ 0.10 &   1\\
8379927	& 120.288$^{+0.017}_{-0.018}$	&	6067 $\pm$ 120   	&  -0.10 $\pm$ 0.15 &   3	\\
8394589	& 109.488$^{+0.034}_{-0.035}$	&	6143 $\pm$ 77		&  -0.29 $\pm$ 0.10 &   1	\\
8424992	& 120.584$^{+0.062}_{-0.064}$	&	5719 $\pm$ 77		&  -0.12 $\pm$ 0.10 &   1      \\
8760414	& 117.230$^{+0.022}_{-0.018}$	&	5873 $\pm$ 77		&  -0.92 $\pm$ 0.10 &   1	\\
9025370	& 132.628$^{+0.030}_{-0.024}$	&	5270 $\pm$ 180  	&  -0.12 $\pm$ 0.18 &   4	\\
9098294	& 108.894$^{+0.023}_{-0.022}$	&	5852 $\pm$ 77		&  -0.18 $\pm$ 0.10 &   1	\\
9139151	& 117.294$^{+0.031}_{-0.032}$	&	6302 $\pm$ 77		&  0.10  $\pm$ 0.10 &   1	\\
9410862	& 107.390$^{+0.050}_{-0.053}$	&	6047 $\pm$ 77		&  -0.31 $\pm$ 0.10 &   1	\\
9955598	& 153.283$^{+0.029}_{-0.032}$	&	5457 $\pm$ 77		&  0.05  $\pm$ 0.10 &   1	\\
9965715	& 97.236$^{+0.041}_{-0.042}$	&	5860 $\pm$ 180  	&  -0.44 $\pm$ 0.18 &	4   \\
10079226 & 116.345$^{+0.059}_{-0.052}$	&	5949 $\pm$ 77		&  0.11 $\pm$ 0.10  &   1	\\
10644253 & 123.080$^{+0.056}_{-0.055}$	&	6045 $\pm$ 77		&  0.06  $\pm$ 0.10 &   1	\\
10963065 & 103.179$^{+0.027}_{-0.027}$	&	6140 $\pm$ 77		&  -0.19 $\pm$ 0.10 &   1	\\
11772920 & 157.746$^{+0.032}_{-0.033}$	&	5180 $\pm$ 180  	&  -0.09 $\pm$ 0.18 &   4   \\
12069424 & 103.277$^{+0.021}_{-0.020}$	&	5825 $\pm$ 50   	&  0.10  $\pm$ 0.03 &   5   \\
12069449 & 116.929$^{+0.012}_{-0.013}$	& 	5750 $\pm$ 50     	&  0.05	 $\pm$ 0.02 &   5   \\
\hline                                   
\end{tabular}
\vspace{0.2cm}\\
Note: All values of the average large frequency separation are adopted from \citet{Lund}. The fourth column  shows the source of the spectroscopic parameters, i.e., (1) \citet{2015Buchhave}, (2) \citet{Casa}, (3) \citet{Pinsonneault}, (4) \citet{Pinso}, and (5) \citet{Ram}.
\end{table}
Table~\ref{1} shows the values of spectroscopic parameters, i.e., metallicity, [Fe/H], and effective temperatures, $T_{\rm eff}$, for each star in our  sample. 
Our stellar sample also includes the asteroseismic binary HD 176465 (\citealt{ 2017Whit,2017EPNsa}) and the Sun, whose average large frequency separation and spectroscopic parameters are shown in Table~\ref{hd4}. 

\begin{table}
\caption{Global asteroseismic and spectroscopic parameters of the asteroseismic binary HD 176465 and the Sun.}
\label{hd4}
\begin{tabular}{c c  cc  }        
\hline \hline 
Star name  &   $\Delta \nu$ ($\mu$ Hz)&	$T_{\mbox{eff}}$ (K)	&  [Fe/H] (dex)	\\
\hline 
HD 176465 A	&  146.79 $\pm$ 0.12	&   5830 $\pm$ 90 		&  -0.30 $\pm$ 0.06   	\\
HD 176465 B	&  155.42 $\pm$ 0.13	&	5740 $\pm$ 90		&   -0.30 $\pm$ 0.06 	\\
Sun	        &  138.8  $\pm$ 0.10       &	5777 $\pm$ 65	& 0.00 $\pm$ 0.05  \\
\hline                                  
\end{tabular}
\vspace{0.2cm}\\
Note: All parameters for the binary HD 176465 are adopted from \citet{2017Whit}. The solar average large frequency separation is obtained from \citet{2013Moss}, while the effective temperature and metallicity are from \cite{Malag}.

\end{table}
\subsection{Stellar models and optimisation}
\label{models}
We constructed three stellar grids (namely; A, B, and C) varying only in the treatment of the initial helium mass fraction ($Y_i$), using MESA version 9793. The evolutionary tracks were evolved from the pre-main sequence (PMS) and stellar models were stored starting from the zero-age main-sequence (ZAMS) which we defined as the region along the evolutionary tracks where the model nuclear luminosity is approximately 99\% of it's total luminosity. Two termination criteria were specified during the grid construction i.e., evolutionary tracks were terminated when:
(i) models reach a stellar age of 16 Gyr. This explains why the evolutionary tracks with stellar masses of 0.7 $\rm M_\odot$, 0.8 $\rm M_\odot$, and 0.9 $\rm M_\odot$ shown in Fig.~\ref{tracks} do not reach the subgiant stage.
(ii) They reach a region along the evolutionary track where log$\rho_c$\,=\,4.5 ($\rho_c$ is the model central density). The model selection along the evolution tracks was based on a variation in the central hydrogen abundance of $\sim$ 0.007.

Table~\ref{tab_grid} provides a summary of different grid constituents. The stellar evolutionary tracks in all the grids were varied in mass, M $\in$ [0.7 -- 1.25] $\rm M_\odot$ in steps of 0.05\,$\rm M_\odot$, Z\,$\in$\,[0.004 -- 0.04] in steps of 0.002, and $\alpha_{\rm mlt}$ $\in$ [1.2 -- 3.0] in steps of 0.2. In grid A, the initial helium abundance, $Y_i$\,$\in$\,[0.22\,--\,0.32] in steps of 0.02.
Diffusion of hydrogen and gravitational settling of heavy elements without radiative acceleration was included in all our stellar grids as indicated in Table~\ref{tab_grid} following the description of \citet{Thoul}. We note that element diffusion has been reported to be an efficient transport process in low-mass stars (e.g. \citealt{Vauclair,Valle2015,2017Higl,Nsambab}). Furthermore, radiative acceleration is
reported to have a negligible impact in stars within the same mass range as the Sun, i.e., below 1.2 $\rm M_\odot$. This is because radiative acceleration is systematically weak compared to gravitational settling or gravity in such stars \citep{1998Turcotte,2017Deal,2018Deal}.

\begin{table}
\centering 
\caption{Stellar grid constituents.} 
\begin{tabular}{ccccc}        
\hline\hline 
Grid name &	Mass ($\rm M_\odot$)	&  Diffusion & Overshoot & $\Delta Y$ $/$ $\Delta Z$\\
\hline
\rowcolor{gray!25}
 A 			& 0.7 -- 1.25	  		&	Yes			&	No	&  ...	    \\
 B 			& 0.7 -- 1.25				&	Yes			&	No	&  1.4	    \\
\rowcolor{gray!25}
 C	    	& 0.7 -- 1.25		    &	Yes			&	No	&  2.0	     \\	
\hline                                  
\end{tabular}
\label{tab_grid}
\end{table}
The general input physics used in all the grids include nuclear reaction rates obtained from JINA REACLIB (Joint Institute for Nuclear Astrophysics Reaction Library; \citealt{Cyburt}) version 2.2 with specific rates for $^{12}\rm C(\alpha,\gamma)^{16}\rm O$ and $^{14}\rm N(p,\gamma)^{15}\rm O$ described by \citet{Kunz} and \citet{Imbriani}, respectively. At high temperatures, OPAL tables \citep{Iglesias} were used to cater for opacities while tables from \citet{Ferguson} were used at lower temperatures. All the grids used the 2005 updated version of the OPAL equation of state \citep{Rogers}. The surface boundary of stellar models was described using the standard Grey-Eddington atmosphere. This integrates the atmosphere structure from the photosphere down to an optical depth of 10$^{-4}$. In all the stellar grids, the surface chemical abundances of \citet{Grevesse1998} with $Z_\odot$ of 0.0169  were used in the conversions of [Fe/H] = $\rm{log} (Z_{surface}/X_{surface})_{\rm{star}}$ -- $\rm{log}       (Z_{\rm{surface}}/X_{\rm{surface}})_{\odot}$.  $X_{\rm{surface}}$ and $Z_{\rm{surface}}$ are the surface hydrogen and heavy element mass fractions, respectively.

The adiabatic oscillation frequencies for the spherical mode degrees, $l$ = 0, 1, 2, and 3 were calculated using GYRE oscillation code \citep{2013Townsend}. Stellar model oscillation frequencies are known to suffer from surface effects which need to be corrected  for before being compared to the observed frequencies (see \citealt{Dziembowski,Dalsgaard,2003Roxburgh}). A number of empirical expressions have been suggested to handle these offsets based on the assumption that they follow a known functional form \citep{2008Kjeldsen,2014Ball,Sonoi} and their impact on the inferred stellar parameters explored \citep{2017Ball,Nsambab,Kinnane,10Christ}. We use the two-term surface correction suggested by \citet{2014Ball} because it was found to yield the least systematic uncertainties on the derived stellar parameters of main-sequence stars. The proposed empirical expression describing the frequency difference ($\delta \nu$) takes the form
\begin{equation}
\delta \nu = I^{-1}\left[ a\left(\frac{\nu}{\nu_{\rm{ac}}} \right)^{-1} + b\left(\frac{\nu}{\nu_{\rm{ac}}} \right)^3  \right]~,
\end{equation}
where $\nu_{\rm{ac}}$ is the acoustic cut-off frequency that scales linearly with $\nu_{\rm{max}}$, which is suggested to scale as the stellar surface gravity, $g$, and effective temperature, ${T_{\rm{eff}}}$, i.e., $\nu_{\rm{ac}} \propto \nu_{\rm{max}} \propto gT^{-1/2}_{\rm{eff}}$ \citep{Brow1991A,1995AKjeld}. $I$ is the mode inertia, $a$ and $b$ are free parameters.

We employ AIMS (Asteroseismic Inference on a Massive Scale; \citealt{2019Rendle}), an optimisation tool based on a Bayesian routine and Markov Chain Monte Carlo (MCMC) approach so as to explore the model parameter space and  find models having parameters comparable to the specified sets of classical and seismic observables. 
It is essential to note that we specified equal weights to both classical and seismic observables, thus the total $\chi_{\rm{total}} ^2$ is expressed as 
\begin{equation}
    \chi_{\rm{total}} ^2 = \left(\frac{\rm N_{\rm{classical}}}{\rm N_{\rm{seismic}}} \right) \chi_{\rm{seismic}} ^2 + \chi_{\rm{classical}} ^2 ~~,
\end{equation}
where $\rm N_{\rm{seismic}}$ is the number of seismic observables, $\rm N_{\rm{classical}}$ is the number of classical observables, 

\noindent
$\chi_{\rm{seismic}} ^2 = \displaystyle\sum_{i}^{N} \left(\frac{\nu_i ^{\rm{(obs)}} - \nu_i ^{\rm{(mod)}}}{\sigma (\nu_i)} \right)^2$, 

\noindent
$\chi_{\rm{classical}} ^2 = \left(\frac{T_{\rm{eff}} ^{\rm(obs)} - T_{\rm{eff}} ^{\rm(mod)}}{\sigma (T_{\rm{eff}})}  \right)^2 + \left(\frac{{\rm{[Fe/H]}}^{\rm(obs)} - {\rm{[Fe/H]}}^{\rm(mod)}}{\sigma {(\rm{[Fe/H]})}}  \right)^2$, and

\noindent
${\nu_i ^{\rm{(obs)}}}$ and ${\nu_i ^{\rm{(mod)}}}$ are the observed and model frequencies, respectively.
Although the weight on each observable should be the same from the statistical point of view, in practice it has been noted that giving similar weights to each observation, thus decreasing the relative impact of the classical constraints in the fit, often leads to results that are biased, possibly due to inaccuracies in the models. This matter has been assessed in the context of a recent hare and hound exercise for PLATO (Cunha et al. in Prep). In this article, we aimed at avoiding that possible bias. While this choice results in larger uncertainties in the parameters estimated from the fit, compared to those found when each observable is given the same weight, the results are expected to be more accurate.

Finally, the stellar parameters and their corresponding uncertainties are obtained as the mean and standard derivation of the posterior distributions. The relative differences ($\Delta \phi / \phi_b$) between stellar parameters from different grids is determined using the expression
\begin{equation}
   \frac{\Delta \phi}{\phi} = \left(\frac{\phi_{a} - \phi_{b}}{\phi_b} \right)~,
\end{equation}
where $\phi$ is any stellar parameter (e.g. mass, $M$, radius, $R$, density, $\rho$, initial helium abundance, $Y_i$, age, $t$, etc), and $\phi_a$ and $\phi_b$ are the inferred stellar parameters from grid $a$ and $b$. 
\section{Results and discussion}
\label{results}
In Sect.~\ref{grid_BC}, we compare the results from grids B and C, considering grid B as the reference grid. We also quantify the associated systematic uncertainties arising from the difference in the treatment of initial helium abundance in both grids. A similar comparison is made in Sect.~\ref{grid_ABC} between grids A and B \& A and C, taking grid A as the reference grid. Sect.~\ref{glitch} contains solar parameters derived from the different grids and a comparison between the surface helium abundances from grid A and those estimated using glitch analysis approach.

\subsection{Comparison between grids B and C}
\label{grid_BC}
\begin{figure*}
	\includegraphics[width=\columnwidth]{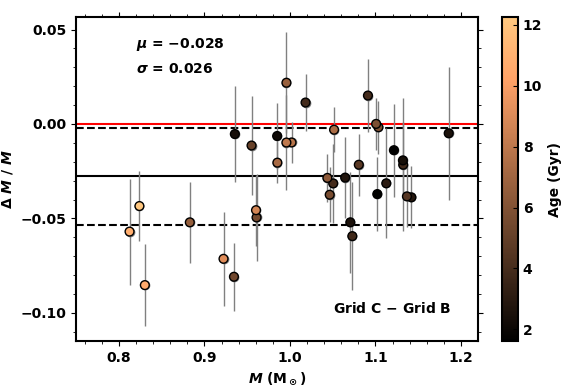}
	    \quad
    \includegraphics[width=\columnwidth]{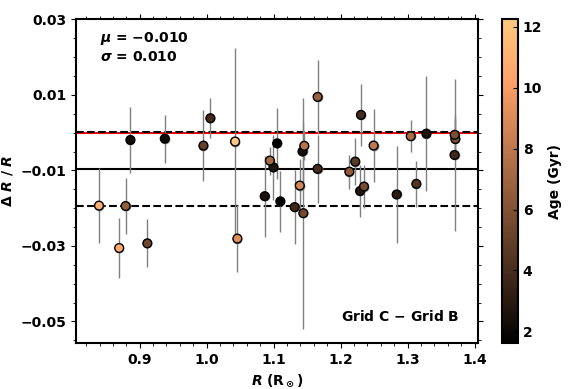}
    	\quad
    \includegraphics[width=\columnwidth]{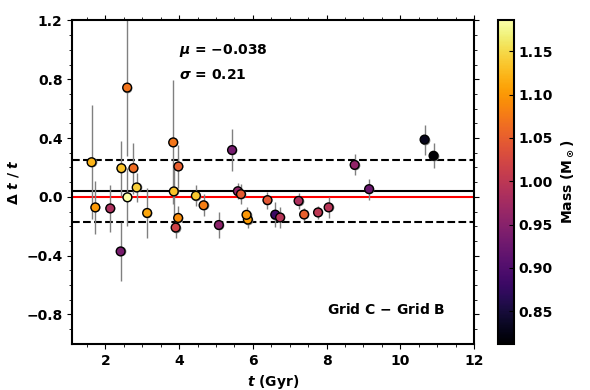}
    	\quad
    \includegraphics[width=\columnwidth]{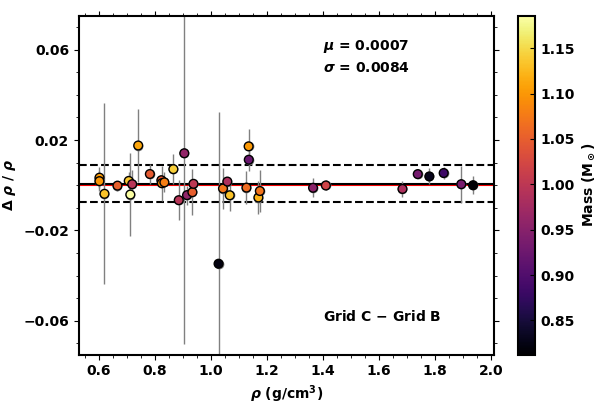}
      \caption{Grid C vs. Grid B: Fractional difference in Mass (top left), radius (top right), age (bottom right) and mean density (bottom right) as a function of stellar parameters from grid B.  The colour-coding is with respect to stellar age (for both top panels) and stellar mass (for both bottom panels). The solid black line indicates the bias ($\mu$), while the scatter ($\sigma$) is represented by the dashed lines. The zero level is represented by the solid red line.}
    \label{MR_BC}
\end{figure*}

We note that grids B and C vary only in the value of the helium-to-heavy element ratio (see Table~\ref{tab_grid}) used in Eq.~(\ref{helium}). The stellar masses from grid C computed with a higher helium-to-heavy element ratio compared to that used in grid B yields lower masses as shown in the top left panel of Fig.~\ref{MR_BC}. This is also illustrated by a bias ($\mu$) $\sim$ $- 0.028$. 
Assuming a fully ionised gas mixture, the mean molecular weight is defined as (e.g.~\citealt{2012Kippe});
\begin{equation}
    \mu_{g} = \frac{4}{6X + Y + 2} ~,
    \label{eq_g}
\end{equation}
and also considering the definition (e.g.~\citealt{2012Kippe});
\begin{equation}
    \mu_{g} = \frac{\bar{m}}{m_H}~,
    \label{eq_g1}
\end{equation}
where $\bar{m}$ is the average mass of the particles (atoms, ions, or molecules) in the gas, $m_H$ is the mass of hydrogen assuming an hydrogen gas.
\begin{figure}
	\includegraphics[width=\columnwidth]{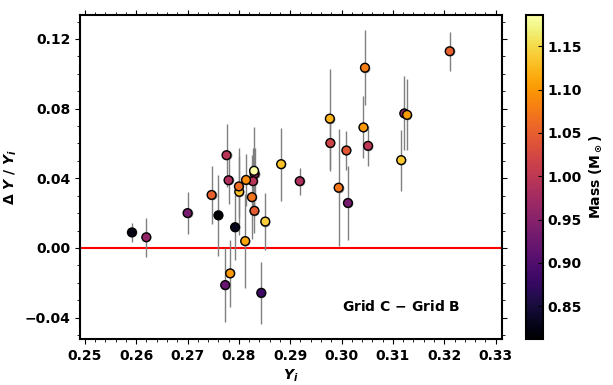}
      \caption{Fractional difference in initial helium abundance as a function of initial helium abundance from grid B. The colour-coding is with respect to stellar mass. The zero level is represented by the solid red line.}
    \label{Y_BC}
\end{figure}

\noindent
For a given set of $Z$ values, larger values of $Y$ in grid C are associated with a larger value of $\Delta Y / \Delta Z$ employed in Eq.~(\ref{helium}) compared to that used in grid B (see Fig.~\ref{Y_BC}). Considering the constraint $X + Y + Z = 1$ and for a fixed $Z$, increase in $Y$ leads to increase in the $\mu_g$ (see Eq.~\ref{eq_g} and Eq.~\ref{eq_g1}), which in turn increases the rate of energy production since more nuclear reactions are taking place per unit time due to the increase in central temperature, i.e., $T_{\rm c}$ $\propto$ $\mu_g$.
This yields higher luminosity and effective temperature. However, in order to prevent this increase in luminosity, which is indirectly constrained using the effective temperature (part of the classical constraints) and seismic data via the Stefan-Boltzmann's relation (e.g. \citealt{1884Boltzman,2015Paul,2018Montambaux}), grid C yields optimal model solutions with lower masses and radii (see both top panels of  Fig.~\ref{MR_BC}) compared to grid B so as to satisfy the required effective temperature and luminosity. The left bottom panel of Fig.~\ref{MR_BC} shows a relatively good agreement between ages from both grids B and C with an expected correlation with mass, i.e.,  high mass stars are younger compared to low mass stars. However, a large scatter of $\sim$ 21\% between the ages can be noted.

We note that both grids B and C yield stellar mean densities which are in good agreement, with a bias of $\sim$ 0.07\% and a scatter of $\sim$ 0.84\% (see bottom right panel of  Fig.~\ref{MR_BC}). This is expected since  we used the same set of individual oscillation frequencies which contain information of the stellar mean density (e.g. \citealt{1986Ulrich,2010Aerts,2013Belkacem,2014Su}).

\subsection{Comparison between grids A and B \& A and C}
\label{grid_ABC}
\begin{figure}
	\includegraphics[width=\columnwidth]{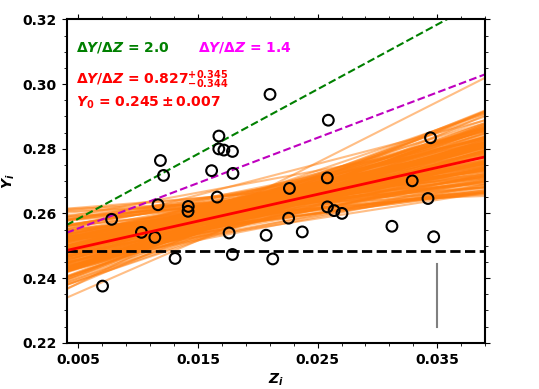}
      \caption{Initial helium abundance as a function of initial heavy elements abundance from grid A which does not constrain the chemical composition with the helium-to-heavy element ratio. The horizontal black dashed line shows the primordial helium abundance from \citet{Cyburt}. The green and magenta dashed lines correspond to the helium-to-heavy element ratios used in grid C and grid B, respectively. The red solid line correspond to a fit obtained using MCMC, characterised by a slope and intercept shown in red on the panel. The orange band around the red solid line corresponds to the Monte Carlo regression uncertainty. The vertical grey line represents the average error on the initial helium abundance values. The median uncertainty on the initial metal mass fraction abundance is 0.002.}
    \label{Y_A}
\end{figure}

We now address the impact of using free  initial helium abundance values (i.e., as carried out in grid A) on the inferred stellar parameters. Fig.~\ref{Y_A} shows a scatter of the preferred initial helium abundance values for the optimal model solutions of our stellar sample. 
It is evident that this scatter does not suggest any relation between initial helium abundance and initial metal abundance. In fact, using MCMC while taking into consideration the associated errors on the initial helium abundance values, we find a slope of $\Delta Y / \Delta Z$\,=\,0.827\,$^{+0.345} _{-0.344}$ and $Y_0 = 0.245 \pm 0.007$.
This $\Delta Y / \Delta Z$ value is lower than those based on observations, i.e., employed in grids B and C (also see Sect.~\ref{intro}). This is consistent with some literature findings, e.g. \citet{Aguirre,Aguirre1}. The $Y_0 = 0.245 \pm 0.007$  is also consistent with that suggested by \citet{Cyburt}, i.e., 0.2484$^{+0.0004}_{-0.0005}$. 
{\it{One would then ask, why do we continue to use Eq.~(\ref{helium}) to determine the model initial helium abundance?}}. This is mainly because observations suggest a $\Delta Y / \Delta Z$ ratio value (e.g. \citealt{Jimenez,2007MCasagrand,Balser}) and also implementing Eq.~(\ref{helium}) reduces the grid parameter space and is therefore computationally less expensive. 
We suggest that a single value of $\Delta Y / \Delta Z$ can be used to describe globally and on average the $Y$ enrichment in relation to $Z$ for a sample of stars, but not from star-to-star (in particularly when describing population I stars). 

\begin{figure*}
	\includegraphics[width=\columnwidth]{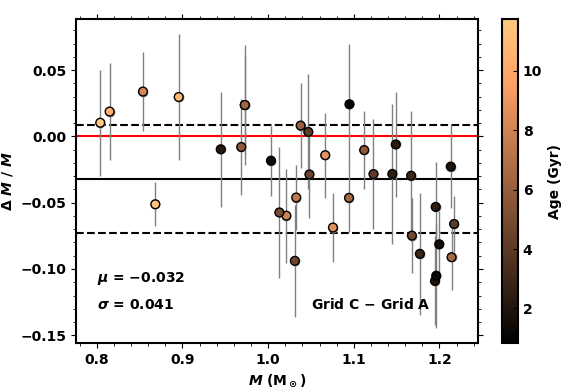}
	    \quad
    \includegraphics[width=\columnwidth]{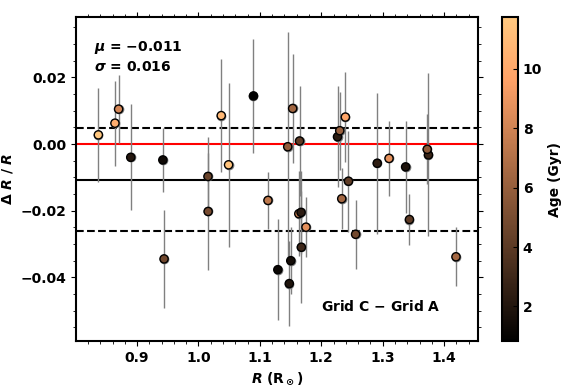}
    	\quad
    \includegraphics[width=\columnwidth]{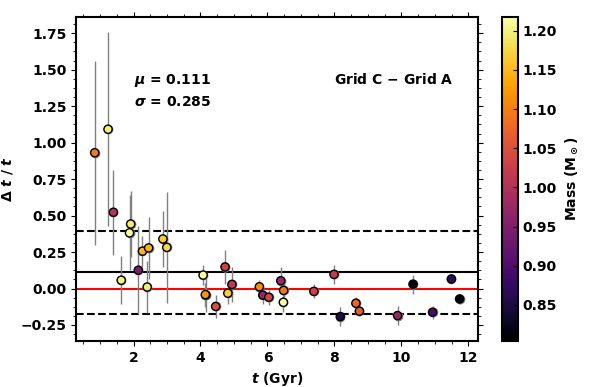}
    	\quad
    \includegraphics[width=\columnwidth]{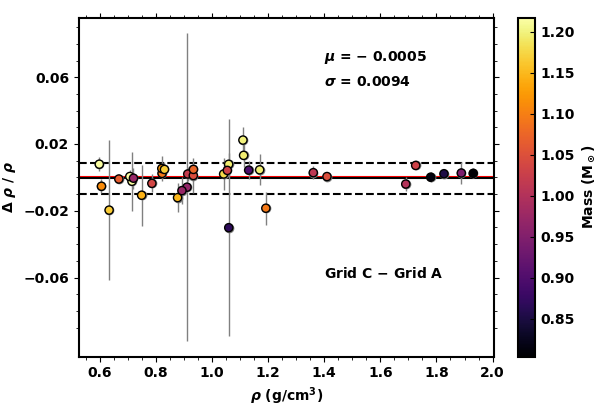}
      \caption{Grid C vs. Grid A: Fractional difference in Mass (top left), radius (top right), age (bottom right) and mean density (bottom right) as a function of stellar parameters from grid A.  The colour-coding is with respect to stellar age (for both top panels) and stellar mass (for both bottom panels). The solid black line indicates the bias ($\mu$), while the scatter ($\sigma$) is represented by the dashed lines. The zero level is represented by the solid red line.}
    \label{MRtD_AC}
\end{figure*}

\begin{figure*}
	\includegraphics[width=\columnwidth]{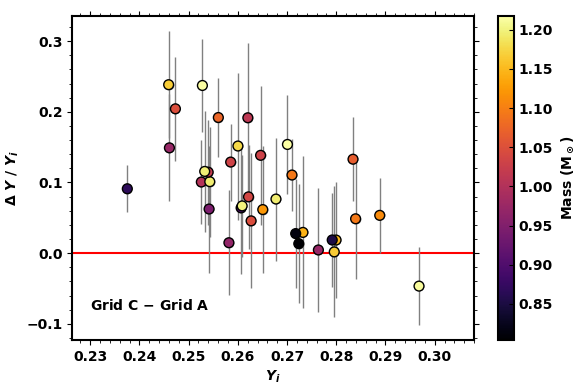}
	\quad
		\includegraphics[width=\columnwidth]{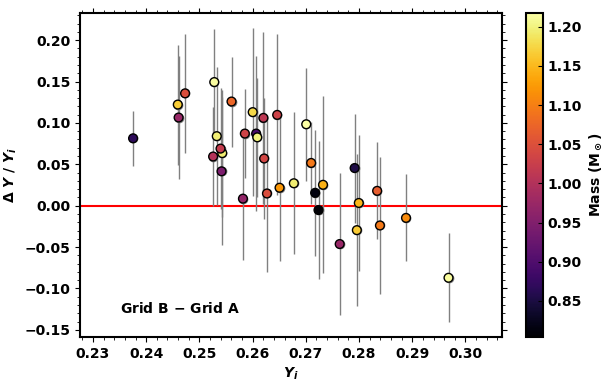}
      \caption{Fractional difference in initial helium abundance as a function of initial helium abundance from A. Left panel: comparison between grid C and grid A. Right panel: comparison between grid B and grid A. The colour-coding is with respect to stellar mass from grid A. The zero level is represented by the solid red line.}
    \label{Y_AC}
\end{figure*}

\begin{figure*}
	\includegraphics[width=\columnwidth]{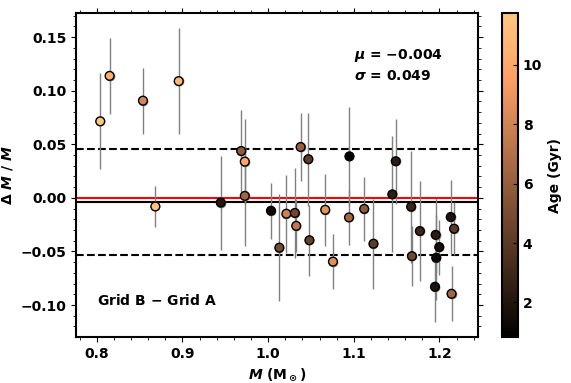}
	    \quad
    \includegraphics[width=\columnwidth]{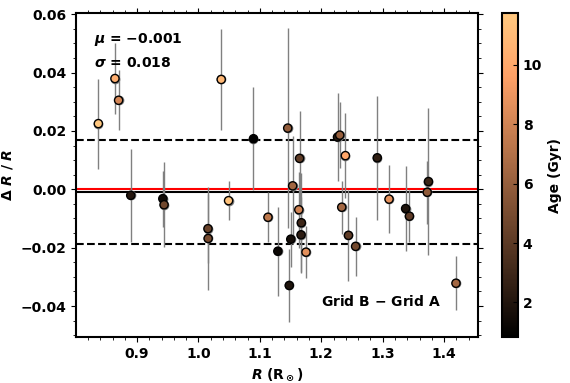}
    	\quad
    \includegraphics[width=\columnwidth]{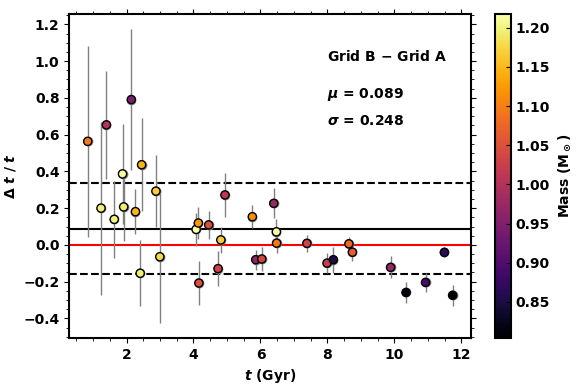}
    	\quad
    \includegraphics[width=\columnwidth]{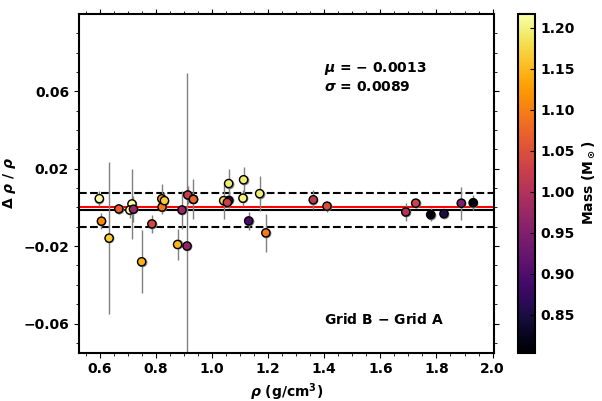}
      \caption{Grid B vs. Grid A: Fractional difference in Mass (top left), radius (top right), age (bottom right) and mean density (bottom right) as a function of stellar parameters from grid A.  The colour-coding is with respect to stellar age (for both top panels) and stellar mass (for both bottom panels). The solid black line indicates the bias ($\mu$), while the scatter ($\sigma$) is represented by the dashed lines. The zero level is represented by the solid red line.}
    \label{MRtD_AB}
\end{figure*}

{\it{In this context, it is important to understand how the inferred stellar parameters differ when grids with a constant $\Delta Y / \Delta Z$ ratio and grids with free initial helium abundance are employed}}.
The top left panels of Fig.~\ref{MRtD_AC} and Fig.~\ref{MRtD_AB} show the relative fractional differences in mass when grid A is compared to grid C and B, respectively. Over all, grid C and B yield lower masses compared to grid A with a scatter of 4.1\% and 4.9\%, and a bias of -3.2\% and -0.4\%, respectively. Furthermore, the top left panel of Fig.~\ref{MRtD_AB} shows that a handful of stars with masses below $\sim$\,0.9 M$_\odot$ from grid B have higher masses compared to those from grid A.
Similar results are seen in radius, i.e., top right panels of Fig.~\ref{MRtD_AC} and Fig.~\ref{MRtD_AB} with a scatter of 1.6\% and 1.8\%, respectively. The systematic uncertainties on mass are consistent with findings of \citet{Lebreton2014} who reported a scatter of $\sim$ 5\% in mass arising from the treatment of initial helium mass fraction when characterising the CoRoT exoplanet-host HD 52265. We note that HD 52265 has a mass slightly higher than that of our stellar sample, i.e., in the range [1.14 - 1.32]\,M$_\odot$.
The left and right panels of Fig.~\ref{Y_AC} show that the optimal models from grid A generally have lower initial helium abundances compared to optimal models from grid C and B, respectively.
This may in part be the origin of the differences observed in mass and radius as discussed in Sect.~\ref{grid_BC}. An excellent agreement in the mean density is shown in the right bottom panels of Fig.~\ref{MRtD_AC} and Fig.~\ref{MRtD_AB} with a scatter of $\sim$ 0.9\%. Despite the low biases in stellar ages (see bottom left panels of of Fig.~\ref{MRtD_AC} and Fig.~\ref{MRtD_AB}), we report significant systematic uncertainties in stellar ages of up-to $\sim$\,29\%. 

\begin{figure*}
	\includegraphics[width=\columnwidth]{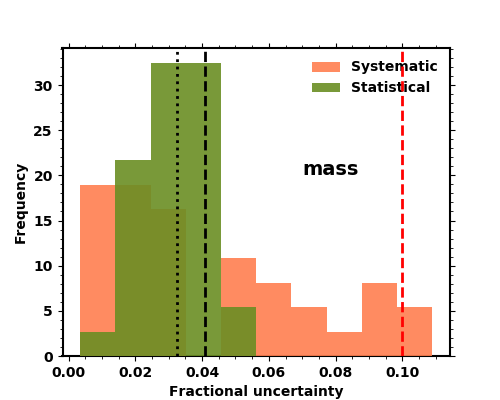}
	    \quad
    \includegraphics[width=\columnwidth]{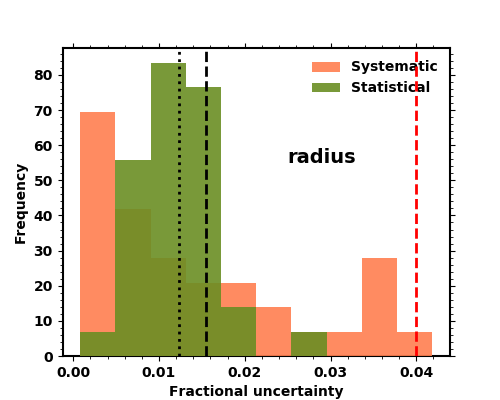}
        \quad
    \includegraphics[width=\columnwidth]{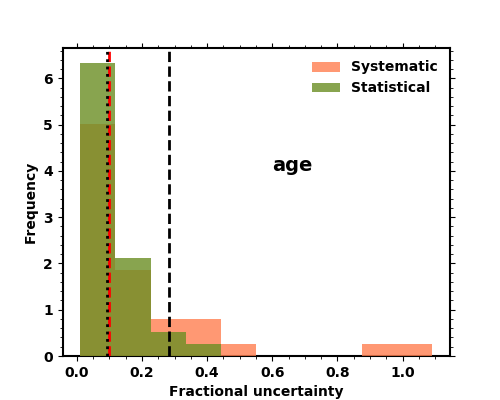}
    	\quad
    \includegraphics[width=\columnwidth]{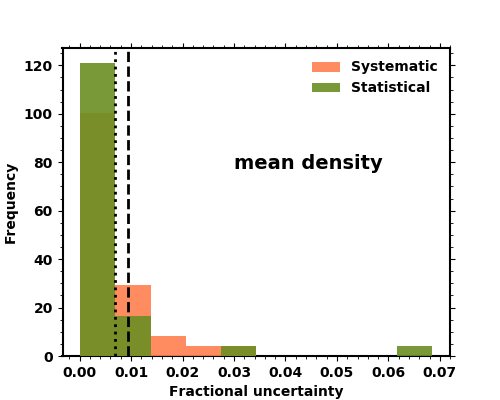}
      \caption{Distributions of statistical uncertainties from grid A (green) and systematic uncertainties from comparing stellar parameters inferred from grids A and C (salmon). Median values are represented by dotted and dashed black lines for the statistical and systematic uncertainty, respectively. The dashed red lines represent the PLATO accuracy requirement limits for stellar parameters (mass, radius and age) of exoplanet-hosts. }
    \label{systematics}
\end{figure*}

To understand how relevant these systematic uncertainties are, we compare them to the statistical uncertainties in Fig.~\ref{systematics}. We note that only a comparison between grids A and C is shown. This is because the systematics between grids A and C are commensurate with those between grids A and B.
However, it is worth noting that significant biases in mass and radius exist as shown in Fig.~\ref{MRtD_AC} and Fig.~\ref{MRtD_AB}. The left bottom panel of Fig.~\ref{systematics} shows that systematic uncertainties in age are greater than statistical uncertainties, highlighting the need for these to be accounted for during model parameter comparisons if a grid with fixed $\Delta Y / \Delta Z$ value is considered.
Fig.~\ref{systematics} also shows a comparison between the systematic uncertainties and the ESA's PLATO accuracy requirements for stellar parameters of exoplanet-host. It is interesting to report that the treatment of initial helium abundance in stellar model grids yields masses and radii within the  expected PLATO accuracy limits, i.e., 2\% - 4\% for the radius and 10\% - 15\% for the mass
of the exoplanet-host \citep{2014Benz}. The left bottom panel of Fig.~\ref{systematics} shows that variations in the treatment of the initial helium abundance yield larger systematic uncertainties on stellar age compared to the required PLATO exoplanet-host accuracy age limits, i.e., 10\% on ages \citep{2014Benz}.

\subsection{Solar parameters and surface helium abundances using acoustic glitches }
\label{glitch}

\begin{figure}
	\includegraphics[width=\columnwidth]{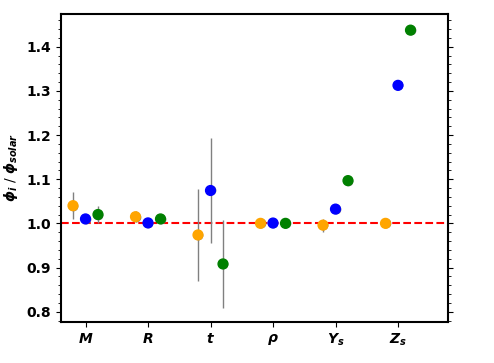}
      \caption{Solar parameters derived using different grids. Parameters from grids A, B, and C are shown in orange, blue, and green, respectively. $\phi_i$ corresponds to the inferred stellar parameter, i.e., mass denoted as $M$, radius denoted as $R$, Age denoted as $t$, mean density denoted as $\rho$, $Z_{\rm s}$ and $Y_{\rm s}$ are the surface metal mass fraction and surface helium mass fraction, respectively. $\phi_{\rm{solar}}$ denotes the solar parameters. See text for details. }
    \label{Sun_plot}
\end{figure}

In order to test how well these different grids reproduce the Sun as a star, we obtained a set of frequencies from \citet{Lund} which are determined from solar data degraded in quality to match that of the {\em {Kepler}} mission. Table~\ref{Sun} shows that grids B and C reproduce the mass and radius of the Sun within 1-$\sigma$ error, while grid A recovers the solar mass and radius within 1.5-$\sigma$. All the grids yield the solar mean density within 1-$\sigma$.
\begin{table*}
\centering 
\caption{{\textbf{ Parameters of the Sun from the different grids}}} 
\begin{tabular}{cccccccc}        
\hline\hline 
Grid	& Mass ($\rm M_{\odot}$) & Radius (\rm $\rm R_{\odot}$) & Age (Gyr)& Density (g/cm$^3$) & $\alpha_{\rm{mlt}}$ & $Y_{\rm{surface}}$  & $Z_{\rm{surface}}$\\
\hline 
\rowcolor{gray!25}
 A 			& 1.04 $\pm$ 0.03 &	1.015 $\pm$ 0.013	& 4.45 $\pm$ 0.25	& 1.410 $\pm$ 0.003 & 1.97  $\pm$ 0.09  &0.247 $\pm$ 0.015 & 0.016 $\pm$ 0.001	    \\
 B 			& 1.01 $\pm$ 0.01 &	1.001 $\pm$ 0.003	& 4.91 $\pm$ 0.31	& 1.411 $\pm$ 0.003& 1.84 $\pm$0.08    & 0.256 $\pm$ 0.004 & 0.021 $\pm$ 0.002   \\
\rowcolor{gray!25}
 C	    & 1.02 $\pm$ 0.02 &	1.010 $\pm$ 0.004	& 4.15 $\pm$ 0.25	& 1.410 $\pm$ 0.002& 1.97 $\pm$ 0.09    & 0.272 $\pm$ 0.005 & 0.023 $\pm$ 0.002    \\
\hline                                   
\end{tabular}
\label{Sun}
\end{table*}
Grids A and B yield the expected solar age within 1-$\sigma$ error but grid C yields a lower age value. Since all our grids employ chemical abundances from \citet{Grevesse1998}, the element mass fraction abundances for this mixture are $X = 0.735$, $Y = 0.248$, and $Z = 0.017$, with $Z/X = 0.023$. These are only satisfied within 1-$\sigma$ error by results from grid A (see Table~\ref{Sun} and Fig.~\ref{Sun_plot}).  We also note that the solar surface helium abundance from grid A is consistent with the helioseismic value. Therefore, we consider stellar parameters derived using grid A to be more accurate than those from grids B and C. The solar reference values considered in Fig.~\ref{Sun_plot} are; mass, $M$\,=\,1 M$_\odot$ (e.g. \citealt{2016Pr}), radius, $R$\,=\,1\,R$_\odot$ \citep{Allen,2012Emilio}, density, $\rho$\,=\,1.410 g/cm$^{-3}$, age, $t$\,=\,4.57 $\pm$ 0.42 Gyr (e.g. \citealt{Connelly}), surface helium mass fraction, $Y_{\rm s}$\,=\,0.248 \citep{2004asu}, and metal mass fraction, $Z_{\rm s}$\,=\,0.017 \citep{Grevesse1998}.

The surface (or envelope) helium abundance ($Y_{\rm{s}}$) can be determined using a ``{\it{semi-direct}}'' approach involving the analysis of a short scale structural variation associated to the helium ionization zones \citep{1988ough,1990Gough}. This is known as the ``helium glitch''. The helium glitch induces a local change in the first adiabatic index, $\Gamma_1$, which creates a variation in the adiabatic sound speed. This in turn creates a signature in the oscillation frequencies and has been observed in the Sun (e.g. \citealt{2005Monte,2007Houdek}), main-sequence stars \citep{2014Mazumd,Verma_2014,20Verma}, and red giant stars (e.g. \citealt{2010Miglio,2014MBroom,2015Vrard,2015Corsa,2020MDr}). Table~\ref{binary} shows a comparison of the surface helium abundances of the binary stars in our sample and those in \citet{Verma_2014} and \citet{2019Verma}. The surface helium abundances from \citet{Verma_2014} and \citet{2019Verma} are based on glitch analysis.
\begin{table*}
\centering 
\caption{Surface helium abundance of two binary stars in our sample. The second and third columns show surface helium abundances from \citet{Verma_2014} and \citet{2019Verma}, respectively, based on glitch analysis and using a set of calibration models (with diffusion) generated using MESA. The fourth column contains results of the best-fitting models from grid A.  } 
\begin{tabular}{ccccccc}        
\hline\hline 
Star	& \citet{Verma_2014} & \citet{2019Verma} & This work\\
\hline 
\rowcolor{gray!25}
 16 Cyg A 			& 0.231 -- 0.251	& 0.232 -- 0.263	& 0.220 -- 0.242	    \\
16 Cyg B			& 0.218 -- 0.266 & 0.245 -- 0.265	& 0.209 -- 0.233    \\
\rowcolor{gray!25}
KIC 6106415			&  --	        & 0.210 -- 0.236	&  0.194 -- 0.224  \\
KIC 6116048	    	& --	            & 0.216 -- 0.238	&   0.192 -- 0.224 \\
\hline                                   
\end{tabular}
\label{binary}
\end{table*}
The surface helium abundance values for the best-fit models of 16 Cyg A, KIC 6106415 and KIC 6116048	from grid A agree within 1-$\sigma$ (see Table~\ref{binary}) with those from \citet{Verma_2014} and \citet{2019Verma}. Our results for 16 Cyg B agree with those of \citet{Verma_2014} and  \citet{2019Verma} within 1-$\sigma$ and 2-$\sigma$, respectively.

\begin{figure}
	\includegraphics[width=\columnwidth]{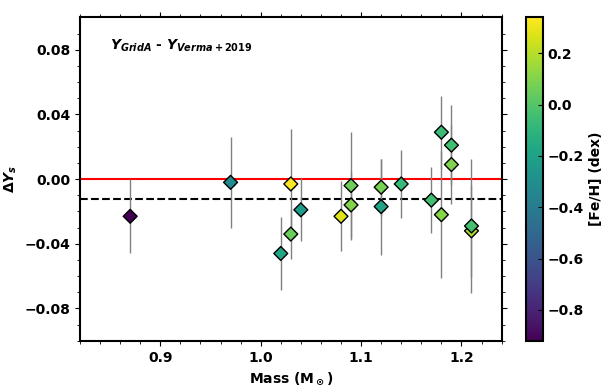}
      \caption{Difference between surface helium abundance values from grid A and those based on glitch analysis from \citet{2019Verma}. The colour-coding is with respect to metallicity. The black dashed line and the solid red line represent the bias ($\sim$ 0.012) and the zero level, respectively. 
    }
    \label{Ydiffere}
\end{figure}

\citet{2019Verma} selected 38 stars from the LEGACY sample, for which the determination of the surface helium abundances using glitch analysis was possible. Of the 38 stars, 19 of them are part of our sample. Fig.~\ref{Ydiffere} shows a comparison between the surface helium abundances of the best-fitting models of grid A for these stars and those determined based on glitch analysis, while employing calibration models generated using MESA in \citet{2019Verma}. The surface helium abundances agree within 1-$\sigma$ to 2-$\sigma$. We note that the glitch analysis approach is not completely model independent since it relies on calibration models (see \citealt{Verma_2014} and \citealt{2019Verma} for details). Therefore, the offset in $Y_s$ of about -0.012 (see Fig.~\ref{Ydiffere}) for most of the stars may be attributed to the differences in the model physics employed in the calibration models used in \citealt{2019Verma} and our grid A models. A vital difference lies in the opacities adopted. The MESA models of \citet{2019Verma} were generated using opacity tables from the Opacity Project (OP) \citep{2005Badnell,2005MSeaton,Ferguson} while our grid A contains models calculated using OPAL opacity tables \citep{Iglesias,Ferguson}. A change in opacities not only creates differences in the depth of convective envelopes but also creates changes in the inferred surface element abundances (e.g. see \citealt{200call}).
The impact of the different model physics employed in stellar grids of solar-type stars on the derived parameters has been addressed in \citet{2004asu,Aguirre,Valle2015,Aguirre1,2017Deal,2018Deal,Nsambaa,2019Nsamb} among others.

\section{Conclusions}
\label{con}
In this article, we have used low-mass ``LEGACY'' sample stars from {\em Kepler} to explore the impact and systematic uncertainties on inferred stellar properties arising from the treatment of initial helium abundance in stellar model grids. Our analysis considers different commonly used approaches for the determination of the initial helium abundance in forward modelling routines. In particular, we have addressed the following questions:

(i) What is the impact of adopting different galactic enrichment ratios on stellar parameters (mainly, mean density, radius, mass and age) inferred using forward modelling routines?

(ii) What kind of systematic uncertainties are induced on the inferred stellar parameters when grids with different galactic enrichment ratios are compared to a grid with helium abundance set as a free parameter?

(iii) How do the statistical uncertainties compare to the systematic uncertainties? Are the systematic uncertainties within the PLATO accuracy requirement limits for exoplanet-host stars?

In sum, our findings indicate that grids constructed with high $\Delta Y / \Delta Z$ yield lower masses and radii compared to grids based on low values of $\Delta Y / \Delta Z$, with systematic uncertainties of $\sim$ 2.6\% and $\sim$ 1.0\% on mass and radius, respectively. The degeneracy between mass and helium is a long standing problem with predictions that it can be broken if precise stellar luminosities are known (see \citealt{Aguirre}). We tested this and found no differences between the results obtained with and without Gaia-based luminosities \citep{2020Moedas}. This is because the precision on the luminosities for the majority of the stars in our sample is not sufficient (i.e., median uncertainty $\sim$ 0.5) to add any strong extra constraint on the optimisation process. 

We also report low initial helium abundance values for the optimal models from the grid with no restrictions set on initial helium abundance, compared to grids which employ an enrichment ratio value  (see Fig.~\ref{Y_AC}). This directly impacts on the inferred stellar masses and radii. 
In addition, we found a handful of stars to have initial helium abundances below the primordial Big Bang nucleosynthesis value (see Fig.~\ref{Y_A}). 
The inference of initial helium abundances below the primordial Big Bang nucleosynthesis value is not uncommon in forward modelling (e.g. \citealt{Bonaca2012,Lebreton2014,Metcalfe2014,Aguirre,Aguirre1}).  \citet{Mathur_2012} argue that this may indicate a  problem with one or more of the observational constraints. This has also been attributed to the choice of solar metallicity mixtures adopted, more specially \citet{Asplund2009} mixtures (see \citealt{Lebreton2014}). \citet{Bonaca2012} demonstrates that in many cases using the solar-calibrated mixing length parameter ($\alpha_{\rm{mlt}}$) would lead to estimates of initial helium abundances that are lower than the primordial helium abundance. In this article, we neither use a solar-calibrated mixing length parameter nor \citet{Asplund2009} mixtures, which may explain why the majority of the solutions for our stellar sample lie within the expected initial helium abundance range (see Fig.~\ref{Y_A}). In fact, given that observations have associated errors, it is expectable that the inferred initially helium abundance may be below the primordial value for a few stars, if stars with low initially helium abundance exist in our sample. In our sample all values of the initial helium abundance are 1-$\sigma$ consistent with a value above the primordial helium abundance. Hence, we find no significant indication for the problem reported by other authors in our study.  

This work gives a detailed insight and quantifies the biases and systematic uncertainties that arise from the treatment of the initial helium abundance in stellar grids. This is important for the ESA's PLATO preparatory work concerning the construction of stellar model grids to be used to infer stellar parameters for stars to be observed by the PLATO mission. The findings in this work are encouraging in a way that the systematic uncertainties on radius and mass found when adopting different approaches for the initial helium abundance determination lie within the required PLATO accuracy limits for stellar parameters of exoplanet-hosts. However, we call attention to the fact that the accuracy limits on the age may not be satisfied.

\section*{Acknowledgements}
The authors acknowledge the dedicated team behind the NASA'S {\it{Kepler}} 
missions.
B.N thanks Verma Kuldeep, Achim Weiss, and the  Stellar Evolution research group at Max-Planck-Institut f\"{u}r Astrophysik (MPA) for the useful comments on this article.
B.N  also acknowledges postdoctoral funding from the Alexander von Humboldt Foundation.
AGH, JCS and BN acknowledge funding support from Spanish public funds (including FEDER funds) for research under projects ESP2017-87676-C5-2-R and ESP2017-87676-C5-5-R. 
T.L.C.~acknowledges support from the European Union's Horizon 2020 research and innovation programme under the Marie Sk\l{}odowska-Curie grant agreement No.~792848 (PULSATION).
J.F. wishes to honor the memory of  Johannes Andersen, recently passed way.  J.F. acknowledges funding from POCH and Portuguese FCT grant SFRH/BSAB/143060/2018. CITEUC is funded by National Funds through FCT - Foundation for Science and Technology (Projects UID-PB/00611/2020).
M. S. Cunha is supported by national funds through Funda\c{c}\~{a}o para a Ci\^{e}ncia e a Tecnologia (FCT, Portugal) - in the form of a work
contract and through the research grants UIDB/04434/2020, UIDP/04434/2020 and PTDC/FIS-AST/30389/2017,
and by FEDER - Fundo Europeu de Desenvolvimento Regional through COMPETE2020 - Programa Operacional Competitividade e Internacionaliz\'{a}cao (grant: POCI-01-0145-FEDER-030389).
AGH also acknowledges support from ``FEDER/Junta de Andaluc\'{\i}a-Consejer\'{\i}a de Econom\'{\i}a y Conocimiento'' under project E-FQM-041-UGR18 by Universidad de Granada. JCS also acknowledges support from project RYC-2012-09913 under the `Ram\'on y Cajal' program of the Spanish Ministry of Science and Education.
J.C. is funded by the Fundamental Research Funds for the Central Universities (grant: 19lgpy278). B.A acknowledges support from the FCT PhD programme PD/BD/135226/2017. We thank the reviewer for the constructive remarks.
\section*{Data availability}
The data underlying this article are available in the article. The details of the ``inlist'' files used in the stellar evolution code (MESA) are described in Sect.~\ref{models} and can easily be reproduced.




\bibliographystyle{mnras}
\bibliography{mnras} 

\begin{thebibliography}{}
\makeatletter
\relax
\def\mn@urlcharsother{\let\do\@makeother \do\$\do\&\do\#\do\^\do\_\do\%\do\~}
\def\mn@doi{\begingroup\mn@urlcharsother \@ifnextchar [ {\mn@doi@}
  {\mn@doi@[]}}
\def\mn@doi@[#1]#2{\def\@tempa{#1}\ifx\@tempa\@empty \href
  {http://dx.doi.org/#2} {doi:#2}\else \href {http://dx.doi.org/#2} {#1}\fi
  \endgroup}
\def\mn@eprint#1#2{\mn@eprint@#1:#2::\@nil}
\def\mn@eprint@arXiv#1{\href {http://arxiv.org/abs/#1} {{\tt arXiv:#1}}}
\def\mn@eprint@dblp#1{\href {http://dblp.uni-trier.de/rec/bibtex/#1.xml}
  {dblp:#1}}
\def\mn@eprint@#1:#2:#3:#4\@nil{\def\@tempa {#1}\def\@tempb {#2}\def\@tempc
  {#3}\ifx \@tempc \@empty \let \@tempc \@tempb \let \@tempb \@tempa \fi \ifx
  \@tempb \@empty \def\@tempb {arXiv}\fi \@ifundefined
  {mn@eprint@\@tempb}{\@tempb:\@tempc}{\expandafter \expandafter \csname
  mn@eprint@\@tempb\endcsname \expandafter{\@tempc}}}

\bibitem[\protect\citeauthoryear{{Adibekyan}, {Sousa}, {Santos}, {Delgado
  Mena}, {Gonz{\'a}lez Hern{\'a}ndez}, {Israelian}, {Mayor}  \&
  {Khachatryan}}{{Adibekyan} et~al.}{2012}]{2012Adib}
{Adibekyan} V.~Z.,  {Sousa} S.~G.,  {Santos} N.~C.,  {Delgado Mena} E.,
  {Gonz{\'a}lez Hern{\'a}ndez} J.~I.,  {Israelian} G.,  {Mayor} M.,
  {Khachatryan} G.,  2012, \mn@doi [A\&A] {10.1051/0004-6361/201219401}, \href
  {https://ui.adsabs.harvard.edu/abs/2012A&A...545A..32A} {545, A32}

\bibitem[\protect\citeauthoryear{{Aerts}, {Christensen-Dalsgaard}  \&
  {Kurtz}}{{Aerts} et~al.}{2010}]{2010Aerts}
{Aerts} C.,  {Christensen-Dalsgaard} J.,   {Kurtz} D.~W.,  2010,
  {Asteroseismology}.
Springer Netherlands

\bibitem[\protect\citeauthoryear{{Allen}}{{Allen}}{1976}]{Allen}
{Allen} C.~W.,  1976, {Astrophysical Quantities}.
3rd.ed. London: AtholonePress

\bibitem[\protect\citeauthoryear{{Anders} \& {Grevesse}}{{Anders} \&
  {Grevesse}}{1989}]{Anders1989}
{Anders} E.,  {Grevesse} N.,  1989, \mn@doi [GeCoA]
  {10.1016/0016-7037(89)90286-X}, \href
  {http://adsabs.harvard.edu/abs/1989GeCoA..53..197A} {53, 197}

\bibitem[\protect\citeauthoryear{{Angelou}, {Bellinger}, {Hekker}, {Mints},
  {Elsworth}, {Basu}  \& {Weiss}}{{Angelou} et~al.}{2020}]{2020Angelou}
{Angelou} G.~C.,  {Bellinger} E.~P.,  {Hekker} S.,  {Mints} A.,  {Elsworth} Y.,
   {Basu} S.,   {Weiss} A.,  2020, \mn@doi [MNRAS] {10.1093/mnras/staa390},
  \href {https://ui.adsabs.harvard.edu/abs/2020MNRAS.493.4987A} {493, 4987}

\bibitem[\protect\citeauthoryear{{Asplund}, {Grevesse}, {Sauval}  \&
  {Scott}}{{Asplund} et~al.}{2009}]{Asplund2009}
{Asplund} M.,  {Grevesse} N.,  {Sauval} A.~J.,   {Scott} P.,  2009, \mn@doi
  [Annual Review of Astron and Astrophys,]
  {10.1146/annurev.astro.46.060407.145222}, \href
  {http://adsabs.harvard.edu/abs/2009ARA%26A..47..481A} {47, 481}

\bibitem[\protect\citeauthoryear{{Badnell}, {Bautista}, {Butler}, {Delahaye},
  {Mendoza}, {Palmeri}, {Zeippen}  \& {Seaton}}{{Badnell}
  et~al.}{2005}]{2005Badnell}
{Badnell} N.~R.,  {Bautista} M.~A.,  {Butler} K.,  {Delahaye} F.,  {Mendoza}
  C.,  {Palmeri} P.,  {Zeippen} C.~J.,   {Seaton} M.~J.,  2005, \mn@doi [MNRAS]
  {10.1111/j.1365-2966.2005.08991.x}, \href
  {https://ui.adsabs.harvard.edu/abs/2005MNRAS.360..458B} {360, 458}

\bibitem[\protect\citeauthoryear{{Bahcall}, {Serenelli}  \&
  {Pinsonneault}}{{Bahcall} et~al.}{2004}]{200call}
{Bahcall} J.~N.,  {Serenelli} A.~M.,   {Pinsonneault} M.,  2004, \mn@doi [ApJ]
  {10.1086/423027}, \href
  {https://ui.adsabs.harvard.edu/abs/2004ApJ...614..464B} {614, 464}

\bibitem[\protect\citeauthoryear{{Ball} \& {Gizon}}{{Ball} \&
  {Gizon}}{2014}]{2014Ball}
{Ball} W.~H.,  {Gizon} L.,  2014, \mn@doi [A\&A] {10.1051/0004-6361/201424325},
  \href {https://ui.adsabs.harvard.edu/abs/2014A&A...568A.123B} {568, A123}

\bibitem[\protect\citeauthoryear{{Ball} \& {Gizon}}{{Ball} \&
  {Gizon}}{2017}]{2017Ball}
{Ball} W.~H.,  {Gizon} L.,  2017, \mn@doi [A\&A] {10.1051/0004-6361/201630260},
  \href {https://ui.adsabs.harvard.edu/abs/2017A&A...600A.128B} {600, A128}

\bibitem[\protect\citeauthoryear{{Balser}}{{Balser}}{2006}]{Balser}
{Balser} D.~S.,  2006, \mn@doi [Astronomical Journal] {10.1086/508515}, \href
  {http://adsabs.harvard.edu/abs/2006AJ....132.2326B} {132, 2326}

\bibitem[\protect\citeauthoryear{{Basu} \& {Antia}}{{Basu} \&
  {Antia}}{2004}]{2004asu}
{Basu} S.,  {Antia} H.~M.,  2004, in {Danesy} D.,  ed.,  ESA Special
  Publication Vol. 559, SOHO 14 Helio- and Asteroseismology: Towards a Golden
  Future. p.~317

\bibitem[\protect\citeauthoryear{Basu \& Kinnane}{Basu \&
  Kinnane}{2018}]{Kinnane}
Basu S.,  Kinnane A.,  2018, \mn@doi [The Astrophysical Journal]
  {10.3847/1538-4357/aae922}, 869, 8

\bibitem[\protect\citeauthoryear{Basu, Chaplin  \& Elsworth}{Basu
  et~al.}{2010}]{Basu_2010}
Basu S.,  Chaplin W.~J.,   Elsworth Y.,  2010, \mn@doi [The Astrophysical
  Journal] {10.1088/0004-637x/710/2/1596}, 710, 1596

\bibitem[\protect\citeauthoryear{{Belkacem}, {Samadi}, {Mosser}, {Goupil}  \&
  {Ludwig}}{{Belkacem} et~al.}{2013}]{2013Belkacem}
{Belkacem} K.,  {Samadi} R.,  {Mosser} B.,  {Goupil} M.~J.,   {Ludwig} H.~G.,
  2013, in {Shibahashi} H.,  {Lynas-Gray} A.~E.,  eds,  Astronomical Society of
  the Pacific Conference Series Vol. 479, Progress in Physics of the Sun and
  Stars: A New Era in Helio- and Asteroseismology. p.~61 (\mn@eprint {arXiv}
  {1307.3132})

\bibitem[\protect\citeauthoryear{{Bellinger}, {Hekker}, {Angelou}, {Stokholm}
  \& {Basu}}{{Bellinger} et~al.}{2019}]{2019Bellinger}
{Bellinger} E.~P.,  {Hekker} S.,  {Angelou} G.~C.,  {Stokholm} A.,   {Basu} S.,
   2019, \mn@doi [A\&A] {10.1051/0004-6361/201834461}, \href
  {https://ui.adsabs.harvard.edu/abs/2019A&A...622A.130B} {622, A130}

\bibitem[\protect\citeauthoryear{{Boltzmann}}{{Boltzmann}}{1884}]{1884Boltzman}
{Boltzmann} L.,  1884, \mn@doi [Annalen der Physik] {10.1002/andp.18842580616},
  \href {https://ui.adsabs.harvard.edu/abs/1884AnP...258..291B} {258, 291}

\bibitem[\protect\citeauthoryear{{Bonaca} et~al.,}{{Bonaca}
  et~al.}{2012}]{Bonaca2012}
{Bonaca} A.,  et~al., 2012, \mn@doi [ApJL] {10.1088/2041-8205/755/1/L12}, \href
  {http://adsabs.harvard.edu/abs/2012ApJ...755L..12B} {755, L12}

\bibitem[\protect\citeauthoryear{{Broomhall} et~al.,}{{Broomhall}
  et~al.}{2014}]{2014MBroom}
{Broomhall} A.~M.,  et~al., 2014, \mn@doi [MNRAS] {10.1093/mnras/stu393}, \href
  {https://ui.adsabs.harvard.edu/abs/2014MNRAS.440.1828B} {440, 1828}

\bibitem[\protect\citeauthoryear{{Brown}, {Gilliland}, {Noyes}  \&
  {Ramsey}}{{Brown} et~al.}{1991}]{Brow1991A}
{Brown} T.~M.,  {Gilliland} R.~L.,  {Noyes} R.~W.,   {Ramsey} L.~W.,  1991,
  \mn@doi [ApJ] {10.1086/169725}, \href
  {https://ui.adsabs.harvard.edu/abs/1991ApJ...368..599B} {368, 599}

\bibitem[\protect\citeauthoryear{{Buchhave} \& {Latham}}{{Buchhave} \&
  {Latham}}{2015}]{2015Buchhave}
{Buchhave} L.~A.,  {Latham} D.~W.,  2015, \mn@doi [ApJ]
  {10.1088/0004-637X/808/2/187}, \href
  {https://ui.adsabs.harvard.edu/abs/2015ApJ...808..187B} {808, 187}

\bibitem[\protect\citeauthoryear{{Casagrande}, {Flynn}, {Portinari}, {Girardi}
  \& {Jimenez}}{{Casagrande} et~al.}{2007}]{2007MCasagrand}
{Casagrande} L.,  {Flynn} C.,  {Portinari} L.,  {Girardi} L.,   {Jimenez} R.,
  2007, \mn@doi [MNRAS] {10.1111/j.1365-2966.2007.12512.x}, \href
  {https://ui.adsabs.harvard.edu/abs/2007MNRAS.382.1516C} {382, 1516}

\bibitem[\protect\citeauthoryear{{Casagrande} et~al.,}{{Casagrande}
  et~al.}{2014}]{Casa}
{Casagrande} L.,  et~al., 2014, \mn@doi [ApJ] {10.1088/0004-637X/787/2/110},
  \href {http://adsabs.harvard.edu/abs/2014ApJ...787..110C} {787, 110}

\bibitem[\protect\citeauthoryear{{Christensen-Dalsgaard} \&
  {Thompson}}{{Christensen-Dalsgaard} \& {Thompson}}{1997}]{Dalsgaard}
{Christensen-Dalsgaard} J.,  {Thompson} M.~J.,  1997, \mn@doi [MNRAS]
  {10.1093/mnras/284.3.527}, \href
  {http://adsabs.harvard.edu/abs/1997MNRAS.284..527C} {284, 527}

\bibitem[\protect\citeauthoryear{{Connelly}, {Bizzarro}, {Krot}, {Nordlund},
  {Wielandt}  \& {Ivanova}}{{Connelly} et~al.}{2012}]{Connelly}
{Connelly} J.~N.,  {Bizzarro} M.,  {Krot} A.~N.,  {Nordlund} {\r{A}}.,
  {Wielandt} D.,   {Ivanova} M.~A.,  2012, \mn@doi [Science]
  {10.1126/science.1226919}, \href
  {https://ui.adsabs.harvard.edu/abs/2012Sci...338..651C} {338, 651}

\bibitem[\protect\citeauthoryear{{Corsaro}, {De Ridder}  \&
  {Garc{\'\i}a}}{{Corsaro} et~al.}{2015}]{2015Corsa}
{Corsaro} E.,  {De Ridder} J.,   {Garc{\'\i}a} R.~A.,  2015, \mn@doi [A\&A]
  {10.1051/0004-6361/201525922}, \href
  {https://ui.adsabs.harvard.edu/abs/2015A&A...578A..76C} {578, A76}

\bibitem[\protect\citeauthoryear{{Cyburt}, {Fields}  \& {Olive}}{{Cyburt}
  et~al.}{2003}]{Cyburt}
{Cyburt} R.~H.,  {Fields} B.~D.,   {Olive} K.~A.,  2003, \mn@doi [Physics
  Letters B] {10.1016/j.physletb.2003.06.026}, \href
  {http://adsabs.harvard.edu/abs/2003PhLB..567..227C} {567, 227}

\bibitem[\protect\citeauthoryear{{Deal}, {Escobar}, {Vauclair}, {Vauclair},
  {Hui-Bon-Hoa}  \& {Richard}}{{Deal} et~al.}{2017}]{2017Deal}
{Deal} M.,  {Escobar} M.~E.,  {Vauclair} S.,  {Vauclair} G.,  {Hui-Bon-Hoa} A.,
    {Richard} O.,  2017, \mn@doi [A\&A] {10.1051/0004-6361/201629825}, \href
  {https://ui.adsabs.harvard.edu/abs/2017A&A...601A.127D} {601, A127}

\bibitem[\protect\citeauthoryear{{Deal}, {Alecian}, {Lebreton}, {Goupil},
  {Marques}, {LeBlanc}, {Morel}  \& {Pichon}}{{Deal} et~al.}{2018}]{2018Deal}
{Deal} M.,  {Alecian} G.,  {Lebreton} Y.,  {Goupil} M.~J.,  {Marques} J.~P.,
  {LeBlanc} F.,  {Morel} P.,   {Pichon} B.,  2018, \mn@doi [A\&a]
  {10.1051/0004-6361/201833361}, \href
  {https://ui.adsabs.harvard.edu/abs/2018A&A...618A..10D} {618, A10}

\bibitem[\protect\citeauthoryear{{Demarque}, {Guenther}, {Li}, {Mazumdar}  \&
  {Straka}}{{Demarque} et~al.}{2008}]{2008ADemarque}
{Demarque} P.,  {Guenther} D.~B.,  {Li} L.~H.,  {Mazumdar} A.,   {Straka}
  C.~W.,  2008, \mn@doi [Ap\&SS] {10.1007/s10509-007-9698-y}, \href
  {https://ui.adsabs.harvard.edu/abs/2008Ap&SS.316...31D} {316, 31}

\bibitem[\protect\citeauthoryear{{Dr{\'e}au}, {Cunha}, {Vrard}  \&
  {Avelino}}{{Dr{\'e}au} et~al.}{2020}]{2020MDr}
{Dr{\'e}au} G.,  {Cunha} M.~S.,  {Vrard} M.,   {Avelino} P.~P.,  2020, \mn@doi
  [MNRAS] {10.1093/mnras/staa1981}, \href
  {https://ui.adsabs.harvard.edu/abs/2020MNRAS.tmp.2101D} {}

\bibitem[\protect\citeauthoryear{{Dziembowski}, {Paterno}  \&
  {Ventura}}{{Dziembowski} et~al.}{1988}]{Dziembowski}
{Dziembowski} W.~A.,  {Paterno} L.,   {Ventura} R.,  1988, A \& A, \href
  {http://adsabs.harvard.edu/abs/1988A%26A...200..213D} {200, 213}

\bibitem[\protect\citeauthoryear{{Emilio}, {Kuhn}, {Bush}  \&
  {Scholl}}{{Emilio} et~al.}{2012}]{2012Emilio}
{Emilio} M.,  {Kuhn} J.~R.,  {Bush} R.~I.,   {Scholl} I.~F.,  2012, \mn@doi
  [ApJ] {10.1088/0004-637X/750/2/135}, \href
  {https://ui.adsabs.harvard.edu/abs/2012ApJ...750..135E} {750, 135}

\bibitem[\protect\citeauthoryear{{Ferguson}, {Alexander}  et~al.}{{Ferguson}
  et~al.}{2005}]{Ferguson}
{Ferguson} J.~W.,  {Alexander} D.~R.,   et~al., 2005, \mn@doi [ApJ]
  {10.1086/428642}, \href {http://adsabs.harvard.edu/abs/2005ApJ...623..585F}
  {623, 585}

\bibitem[\protect\citeauthoryear{{Frandsen} et~al.,}{{Frandsen}
  et~al.}{2018}]{2018Frandsen}
{Frandsen} S.,  et~al., 2018, \mn@doi [A\&A] {10.1051/0004-6361/201730816},
  \href {https://ui.adsabs.harvard.edu/abs/2018A&A...613A..53F} {613, A53}

\bibitem[\protect\citeauthoryear{Gai, Basu, Chaplin  \& Elsworth}{Gai
  et~al.}{2011}]{Gai_2011}
Gai N.,  Basu S.,  Chaplin W.~J.,   Elsworth Y.,  2011, \mn@doi [The
  Astrophysical Journal] {10.1088/0004-637x/730/2/63}, 730, 63

\bibitem[\protect\citeauthoryear{{Gough}}{{Gough}}{1990}]{1990Gough}
{Gough} D.~O.,  1990, {Comments on Helioseismic Inference}.
{Osaki}, Yoji and {Shibahashi}, Hiromoto, p.~283,
  \mn@doi{10.1007/3-540-53091-6}

\bibitem[\protect\citeauthoryear{{Gough} \& {Thompson}}{{Gough} \&
  {Thompson}}{1988}]{1988ough}
{Gough} D.~O.,  {Thompson} M.~J.,  1988, in {Christensen-Dalsgaard} J.,
  {Frandsen} S.,  eds,  IAU Symposium Vol. 123, Advances in Helio- and
  Asteroseismology. p.~155

\bibitem[\protect\citeauthoryear{{Grevesse} \& {Sauval}}{{Grevesse} \&
  {Sauval}}{1998}]{Grevesse1998}
{Grevesse} N.,  {Sauval} A.~J.,  1998, \mn@doi [Space Science Reviews]
  {10.1023/A:1005161325181}, \href
  {http://adsabs.harvard.edu/abs/1998SSRv...85..161G} {85, 161}

\bibitem[\protect\citeauthoryear{{Grevesse}, {Asplund}  \& {Sauval}}{{Grevesse}
  et~al.}{2007}]{Grevesse2007}
{Grevesse} N.,  {Asplund} M.,   {Sauval} A.~J.,  2007, \mn@doi [Space Science
  Reviews] {10.1007/s11214-007-9173-7}, \href
  {http://adsabs.harvard.edu/abs/2007SSRv..130..105G} {130, 105}

\bibitem[\protect\citeauthoryear{{Higl} \& {Weiss}}{{Higl} \&
  {Weiss}}{2017}]{2017Higl}
{Higl} J.,  {Weiss} A.,  2017, \mn@doi [A\&A] {10.1051/0004-6361/201731008},
  \href {https://ui.adsabs.harvard.edu/abs/2017A&A...608A..62H} {608, A62}

\bibitem[\protect\citeauthoryear{{Houdek} \& {Gough}}{{Houdek} \&
  {Gough}}{2007}]{2007Houdek}
{Houdek} G.,  {Gough} D.~O.,  2007, \mn@doi [MNRAS]
  {10.1111/j.1365-2966.2006.11325.x}, \href
  {https://ui.adsabs.harvard.edu/abs/2007MNRAS.375..861H} {375, 861}

\bibitem[\protect\citeauthoryear{{Iglesias} \& {Rogers}}{{Iglesias} \&
  {Rogers}}{1996}]{Iglesias}
{Iglesias} C.~A.,  {Rogers} F.~J.,  1996, \mn@doi [ApJ] {10.1086/177381}, \href
  {http://adsabs.harvard.edu/abs/1996ApJ...464..943I} {464, 943}

\bibitem[\protect\citeauthoryear{{Imbriani}, {Costantini}  et~al.}{{Imbriani}
  et~al.}{2005}]{Imbriani}
{Imbriani} G.,  {Costantini} H.,   et~al., 2005, \mn@doi [EPJ A]
  {10.1140/epja/i2005-10138-7}, \href
  {http://adsabs.harvard.edu/abs/2005EPJA...25..455I} {25, 455}

\bibitem[\protect\citeauthoryear{{Jiang} et~al.,}{{Jiang}
  et~al.}{2020}]{2020Jiang}
{Jiang} C.,  et~al., 2020, \mn@doi [ApJ] {10.3847/1538-4357/ab8f29}, \href
  {https://ui.adsabs.harvard.edu/abs/2020ApJ...896...65J} {896, 65}

\bibitem[\protect\citeauthoryear{{Jimenez}, {Flynn}, {MacDonald}  \&
  {Gibson}}{{Jimenez} et~al.}{2003}]{Jimenez}
{Jimenez} R.,  {Flynn} C.,  {MacDonald} J.,   {Gibson} B.~K.,  2003, \mn@doi
  [Science] {10.1126/science.1080866}, \href
  {http://adsabs.harvard.edu/abs/2003Sci...299.1552J} {299, 1552}

\bibitem[\protect\citeauthoryear{J{\o}rgensen et~al.,}{J{\o}rgensen
  et~al.}{2020}]{10Christ}
J{\o}rgensen A. C.~S.,  et~al., 2020, \mn@doi [Monthly Notices of the Royal
  Astronomical Society] {10.1093/mnras/staa1480}, 495, 4965

\bibitem[\protect\citeauthoryear{{Kippenhahn}, {Weigert}  \&
  {Weiss}}{{Kippenhahn} et~al.}{2012}]{2012Kippe}
{Kippenhahn} R.,  {Weigert} A.,   {Weiss} A.,  2012, {Stellar Structure and
  Evolution}.
Springer, Berlin, Heidelberg, \mn@doi{10.1007/978-3-642-30304-3}

\bibitem[\protect\citeauthoryear{{Kjeldsen} \& {Bedding}}{{Kjeldsen} \&
  {Bedding}}{1995}]{1995AKjeld}
{Kjeldsen} H.,  {Bedding} T.~R.,  1995, A\&A, \href
  {https://ui.adsabs.harvard.edu/abs/1995A&A...293...87K} {293, 87}

\bibitem[\protect\citeauthoryear{{Kjeldsen}, {Bedding}  \&
  {Christensen-Dalsgaard}}{{Kjeldsen} et~al.}{2008}]{2008Kjeldsen}
{Kjeldsen} H.,  {Bedding} T.~R.,   {Christensen-Dalsgaard} J.,  2008, \mn@doi
  [ApJL] {10.1086/591667}, \href
  {https://ui.adsabs.harvard.edu/abs/2008ApJ...683L.175K} {683, L175}

\bibitem[\protect\citeauthoryear{{Kunz}, {Fey}  et~al.}{{Kunz}
  et~al.}{2002}]{Kunz}
{Kunz} R.,  {Fey} M.,   et~al., 2002, \mn@doi [ApJ] {10.1086/338384}, \href
  {http://adsabs.harvard.edu/abs/2002ApJ...567..643K} {567, 643}

\bibitem[\protect\citeauthoryear{{Lebreton} \& {Goupil}}{{Lebreton} \&
  {Goupil}}{2014}]{Lebreton2014}
{Lebreton} Y.,  {Goupil} M.~J.,  2014, \mn@doi [A \& A]
  {10.1051/0004-6361/201423797}, \href
  {http://adsabs.harvard.edu/abs/2014A%26A...569A..21L} {569, A21}

\bibitem[\protect\citeauthoryear{{Li}, {Bedding}, {Christensen-Dalsgaard},
  {Stello}, {Li}  \& {Keen}}{{Li} et~al.}{2020}]{Li}
{Li} T.,  {Bedding} T.~R.,  {Christensen-Dalsgaard} J.,  {Stello} D.,  {Li} Y.,
    {Keen} M.~A.,  2020, \mn@doi [MNRAS] {10.1093/mnras/staa1350}, \href
  {https://ui.adsabs.harvard.edu/abs/2020MNRAS.495.3431L} {495, 3431}

\bibitem[\protect\citeauthoryear{{Lodders}}{{Lodders}}{2003}]{Lodders2003}
{Lodders} K.,  2003, \mn@doi [ApJ] {10.1086/375492}, \href
  {http://adsabs.harvard.edu/abs/2003ApJ...591.1220L} {591, 1220}

\bibitem[\protect\citeauthoryear{{Lund} et~al.,}{{Lund} et~al.}{2017}]{Lund}
{Lund} M.~N.,  et~al., 2017, \mn@doi [ApJ] {10.3847/1538-4357/835/2/172}, \href
  {http://adsabs.harvard.edu/abs/2017ApJ...835..172L} {835, 172}

\bibitem[\protect\citeauthoryear{{Malagnini} \& {Morossi}}{{Malagnini} \&
  {Morossi}}{1997}]{Malag}
{Malagnini} M.~L.,  {Morossi} C.,  1997, A\&A, \href
  {https://ui.adsabs.harvard.edu/abs/1997A&A...326..736M} {326, 736}

\bibitem[\protect\citeauthoryear{Mathur et~al.,}{Mathur
  et~al.}{2012}]{Mathur_2012}
Mathur S.,  et~al., 2012, \mn@doi [The Astrophysical Journal]
  {10.1088/0004-637x/749/2/152}, 749, 152

\bibitem[\protect\citeauthoryear{{Mazumdar} et~al.,}{{Mazumdar}
  et~al.}{2014}]{2014Mazumd}
{Mazumdar} A.,  et~al., 2014, \mn@doi [ApJ] {10.1088/0004-637X/782/1/18}, \href
  {https://ui.adsabs.harvard.edu/abs/2014ApJ...782...18M} {782, 18}

\bibitem[\protect\citeauthoryear{Metcalfe et~al.,}{Metcalfe
  et~al.}{2010}]{Metcalfe_2010}
Metcalfe T.~S.,  et~al., 2010, \mn@doi [The Astrophysical Journal]
  {10.1088/0004-637x/723/2/1583}, 723, 1583

\bibitem[\protect\citeauthoryear{{Metcalfe} et~al.,}{{Metcalfe}
  et~al.}{2014}]{Metcalfe2014}
{Metcalfe} T.~S.,  et~al., 2014, \mn@doi [ApJS] {10.1088/0067-0049/214/2/27},
  \href {http://adsabs.harvard.edu/abs/2014ApJS..214...27M} {214, 27}

\bibitem[\protect\citeauthoryear{{Miglio} et~al.,}{{Miglio}
  et~al.}{2010}]{2010Miglio}
{Miglio} A.,  et~al., 2010, \mn@doi [A\&A] {10.1051/0004-6361/201015442}, \href
  {https://ui.adsabs.harvard.edu/abs/2010A&A...520L...6M} {520, L6}

\bibitem[\protect\citeauthoryear{{Moedas}, {Nsamba}  \& {Clara}}{{Moedas}
  et~al.}{2020}]{2020Moedas}
{Moedas} N.,  {Nsamba} B.,   {Clara} M.~T.,  2020, arXiv e-prints, \href
  {https://ui.adsabs.harvard.edu/abs/2020arXiv200703362M} {p. arXiv:2007.03362}

\bibitem[\protect\citeauthoryear{{Montambaux}}{{Montambaux}}{2018}]{2018Montambaux}
{Montambaux} G.,  2018, \mn@doi [Foundations of Physics]
  {10.1007/s10701-018-0153-4}, \href
  {https://ui.adsabs.harvard.edu/abs/2018FoPh..tmp...27M} {}

\bibitem[\protect\citeauthoryear{{Monteiro} \& {Thompson}}{{Monteiro} \&
  {Thompson}}{2005}]{2005Monte}
{Monteiro} M. J.~P.~F.~G.,  {Thompson} M.~J.,  2005, \mn@doi [MNRAS]
  {10.1111/j.1365-2966.2005.09246.x}, \href
  {https://ui.adsabs.harvard.edu/abs/2005MNRAS.361.1187M} {361, 1187}

\bibitem[\protect\citeauthoryear{{Mosser} et~al.,}{{Mosser}
  et~al.}{2013}]{2013Moss}
{Mosser} B.,  et~al., 2013, \mn@doi [A\&A] {10.1051/0004-6361/201220435}, \href
  {https://ui.adsabs.harvard.edu/abs/2013A&A...550A.126M} {550, A126}

\bibitem[\protect\citeauthoryear{{Nsamba}, {Monteiro}, {Campante}, {Reese},
  {White}, {Garc{\'\i}a Hern{\'a}ndez}  \& {Jiang}}{{Nsamba}
  et~al.}{2017}]{2017EPNsa}
{Nsamba} B.,  {Monteiro} M.~J.~P.~F.~G.,  {Campante} T.~L.,  {Reese} D.~R.,
  {White} T.~R.,  {Garc{\'\i}a Hern{\'a}ndez} A.,   {Jiang} C.,  2017, in
  European Physical Journal Web of Conferences. p. 05010 (\mn@eprint {arXiv}
  {1611.05698}), \mn@doi{10.1051/epjconf/201716005010}

\bibitem[\protect\citeauthoryear{{Nsamba}, {Campante}, {Monteiro}, {Cunha},
  {Rendle}, {Reese}  \& {Verma}}{{Nsamba} et~al.}{2018a}]{Nsambab}
{Nsamba} B.,  {Campante} T.~L.,  {Monteiro} M.~J.~P.~F.~G.,  {Cunha} M.~S.,
  {Rendle} B.~M.,  {Reese} D.~R.,   {Verma} K.,  2018a, \mn@doi [MNRAS]
  {10.1093/mnras/sty948}, \href
  {http://adsabs.harvard.edu/abs/2018MNRAS.477.5052N} {477, 5052}

\bibitem[\protect\citeauthoryear{{Nsamba}, {Monteiro}, {Campante}, {Cunha}  \&
  {Sousa}}{{Nsamba} et~al.}{2018b}]{Nsambaa}
{Nsamba} B.,  {Monteiro} M.~J.~P.~F.~G.,  {Campante} T.~L.,  {Cunha} M.~S.,
  {Sousa} S.~G.,  2018b, \mn@doi [MNRAS] {10.1093/mnrasl/sly092}, \href
  {http://adsabs.harvard.edu/abs/2018MNRAS.479L..55N} {479, L55}

\bibitem[\protect\citeauthoryear{{Nsamba}, {Campante}, {Monteiro}, {Cunha}  \&
  {Sousa}}{{Nsamba} et~al.}{2019}]{2019Nsamb}
{Nsamba} B.,  {Campante} T.~L.,  {Monteiro} M. J.~P.~F.~G.,  {Cunha} M.~S.,
  {Sousa} S.~G.,  2019, \mn@doi [Frontiers in Astronomy and Space Sciences]
  {10.3389/fspas.2019.00025}, \href
  {https://ui.adsabs.harvard.edu/abs/2019FrASS...6...25N} {6, 25}

\bibitem[\protect\citeauthoryear{{Paul}, {Greenberger}, {Stenholm}  \&
  {Schleich}}{{Paul} et~al.}{2015}]{2015Paul}
{Paul} H.,  {Greenberger} D.~M.,  {Stenholm} S.~T.,   {Schleich} W.~P.,  2015,
  \mn@doi [Physica Scripta Volume T] {10.1088/0031-8949/2015/T165/014027},
  \href {https://ui.adsabs.harvard.edu/abs/2015PhST..165a4027P} {165, 014027}

\bibitem[\protect\citeauthoryear{{Paxton} et~al.,}{{Paxton}
  et~al.}{2013}]{Pax2}
{Paxton} B.,  et~al., 2013, \mn@doi [ApJS] {10.1088/0067-0049/208/1/4}, \href
  {http://adsabs.harvard.edu/abs/2013ApJS..208....4P} {208, 4}

\bibitem[\protect\citeauthoryear{{Paxton} et~al.,}{{Paxton}
  et~al.}{2015}]{Pax3}
{Paxton} B.,  et~al., 2015, \mn@doi [ApJS] {10.1088/0067-0049/220/1/15}, \href
  {http://adsabs.harvard.edu/abs/2015ApJS..220...15P} {220, 15}

\bibitem[\protect\citeauthoryear{{Paxton} et~al.,}{{Paxton}
  et~al.}{2018}]{Pax4}
{Paxton} B.,  et~al., 2018, \mn@doi [ApJS] {10.3847/1538-4365/aaa5a8}, \href
  {http://adsabs.harvard.edu/abs/2018ApJS..234...34P} {234, 34}

\bibitem[\protect\citeauthoryear{{Pinsonneault}, {An}, {Molenda-{\.Z}akowicz},
  {Chaplin}, {Metcalfe}  \& {Bruntt}}{{Pinsonneault}
  et~al.}{2012}]{Pinsonneault}
{Pinsonneault} M.~H.,  {An} D.,  {Molenda-{\.Z}akowicz} J.,  {Chaplin} W.~J.,
  {Metcalfe} T.~S.,   {Bruntt} H.,  2012, \mn@doi [ApJS]
  {10.1088/0067-0049/199/2/30}, \href
  {http://adsabs.harvard.edu/abs/2012ApJS..199...30P} {199, 30}

\bibitem[\protect\citeauthoryear{{Pinsonneault} et~al.,}{{Pinsonneault}
  et~al.}{2014}]{Pinso}
{Pinsonneault} M.~H.,  et~al., 2014, \mn@doi [ApJS]
  {10.1088/0067-0049/215/2/19}, \href
  {http://adsabs.harvard.edu/abs/2014ApJS..215...19P} {215, 19}

\bibitem[\protect\citeauthoryear{{Pr{\v{s}}a} et~al.,}{{Pr{\v{s}}a}
  et~al.}{2016}]{2016Pr}
{Pr{\v{s}}a} A.,  et~al., 2016, \mn@doi [AJ] {10.3847/0004-6256/152/2/41},
  \href {https://ui.adsabs.harvard.edu/abs/2016AJ....152...41P} {152, 41}

\bibitem[\protect\citeauthoryear{{Ram{\'{\i}}rez}, {Mel{\'e}ndez}  \&
  {Asplund}}{{Ram{\'{\i}}rez} et~al.}{2009}]{Ram}
{Ram{\'{\i}}rez} I.,  {Mel{\'e}ndez} J.,   {Asplund} M.,  2009, \mn@doi [A \&
  A] {10.1051/0004-6361/200913038}, \href
  {http://adsabs.harvard.edu/abs/2009A%26A...508L..17R} {508, L17}

\bibitem[\protect\citeauthoryear{{Rauer} et~al.,}{{Rauer}
  et~al.}{2014}]{2014Benz}
{Rauer} H.,  et~al., 2014, \mn@doi [Experimental Astronomy]
  {10.1007/s10686-014-9383-4}, \href
  {https://ui.adsabs.harvard.edu/abs/2014ExA....38..249R} {38, 249}

\bibitem[\protect\citeauthoryear{{Rendle} et~al.,}{{Rendle}
  et~al.}{2019}]{2019Rendle}
{Rendle} B.~M.,  et~al., 2019, \mn@doi [MNRAS] {10.1093/mnras/stz031}, \href
  {https://ui.adsabs.harvard.edu/abs/2019MNRAS.484..771R} {484, 771}

\bibitem[\protect\citeauthoryear{Rodrigues et~al.,}{Rodrigues
  et~al.}{2017}]{Rodrigues}
Rodrigues T.~S.,  et~al., 2017, \mn@doi [Monthly Notices of the Royal
  Astronomical Society] {10.1093/mnras/stx120}, 467, 1433

\bibitem[\protect\citeauthoryear{{Rogers} \& {Nayfonov}}{{Rogers} \&
  {Nayfonov}}{2002}]{Rogers}
{Rogers} F.~J.,  {Nayfonov} A.,  2002, \mn@doi [ApJ] {10.1086/341894}, \href
  {http://adsabs.harvard.edu/abs/2002ApJ...576.1064R} {576, 1064}

\bibitem[\protect\citeauthoryear{{Roxburgh} \& {Vorontsov}}{{Roxburgh} \&
  {Vorontsov}}{2003}]{2003Roxburgh}
{Roxburgh} I.~W.,  {Vorontsov} S.~V.,  2003, \mn@doi [A\&A]
  {10.1051/0004-6361:20031318}, \href
  {https://ui.adsabs.harvard.edu/abs/2003A&A...411..215R} {411, 215}

\bibitem[\protect\citeauthoryear{{Seaton}}{{Seaton}}{2005}]{2005MSeaton}
{Seaton} M.~J.,  2005, \mn@doi [MNRAS] {10.1111/j.1365-2966.2005.00019.x},
  \href {https://ui.adsabs.harvard.edu/abs/2005MNRAS.362L...1S} {362, L1}

\bibitem[\protect\citeauthoryear{{Serenelli} \& {Basu}}{{Serenelli} \&
  {Basu}}{2010}]{Serenelli}
{Serenelli} A.~M.,  {Basu} S.,  2010, \mn@doi [ApJ]
  {10.1088/0004-637X/719/1/865}, \href
  {http://adsabs.harvard.edu/abs/2010ApJ...719..865S} {719, 865}

\bibitem[\protect\citeauthoryear{{Serenelli} et~al.,}{{Serenelli}
  et~al.}{2020}]{2020arSerenelli}
{Serenelli} A.,  et~al., 2020, arXiv e-prints, \href
  {https://ui.adsabs.harvard.edu/abs/2020arXiv200610868S} {p. arXiv:2006.10868}

\bibitem[\protect\citeauthoryear{{Silva Aguirre} et~al.,}{{Silva Aguirre}
  et~al.}{2015}]{Aguirre}
{Silva Aguirre} V.,  et~al., 2015, \mn@doi [MNRAS] {10.1093/mnras/stv1388},
  \href {http://adsabs.harvard.edu/abs/2015MNRAS.452.2127S} {452, 2127}

\bibitem[\protect\citeauthoryear{{Silva Aguirre} et~al.,}{{Silva Aguirre}
  et~al.}{2017}]{Aguirre1}
{Silva Aguirre} V.,  et~al., 2017, \mn@doi [ApJ] {10.3847/1538-4357/835/2/173},
  \href {http://adsabs.harvard.edu/abs/2017ApJ...835..173S} {835, 173}

\bibitem[\protect\citeauthoryear{{Sonoi}, {Samadi}, {Belkacem}, {Ludwig},
  {Caffau}  \& {Mosser}}{{Sonoi} et~al.}{2015}]{Sonoi}
{Sonoi} T.,  {Samadi} R.,  {Belkacem} K.,  {Ludwig} H.~G.,  {Caffau} E.,
  {Mosser} B.,  2015, \mn@doi [A\&A] {10.1051/0004-6361/201526838}, \href
  {https://ui.adsabs.harvard.edu/abs/2015A&A...583A.112S} {583, A112}

\bibitem[\protect\citeauthoryear{{Su{\'a}rez}, {Garc{\'\i}a Hern{\'a}ndez},
  {Moya}, {Rodrigo}, {Solano}, {Garrido}  \& {Rod{\'o}n}}{{Su{\'a}rez}
  et~al.}{2014}]{2014Su}
{Su{\'a}rez} J.~C.,  {Garc{\'\i}a Hern{\'a}ndez} A.,  {Moya} A.,  {Rodrigo} C.,
   {Solano} E.,  {Garrido} R.,   {Rod{\'o}n} J.~R.,  2014, \mn@doi [A\&A]
  {10.1051/0004-6361/201322270}, \href
  {https://ui.adsabs.harvard.edu/abs/2014A&A...563A...7S} {563, A7}

\bibitem[\protect\citeauthoryear{{Th{\'e}ado}, {Vauclair}, {Castro},
  {Charpinet}  \& {Dolez}}{{Th{\'e}ado} et~al.}{2005}]{Vauclair}
{Th{\'e}ado} S.,  {Vauclair} S.,  {Castro} M.,  {Charpinet} S.,   {Dolez} N.,
  2005, \mn@doi [A\&A] {10.1051/0004-6361:20042328}, \href
  {http://adsabs.harvard.edu/abs/2005A%26A...437..553T} {437, 553}

\bibitem[\protect\citeauthoryear{{Thoul}, {Bahcall}  \& {Loeb}}{{Thoul}
  et~al.}{1994}]{Thoul}
{Thoul} A.~A.,  {Bahcall} J.~N.,   {Loeb} A.,  1994, \mn@doi [ApJ]
  {10.1086/173695}, \href {http://adsabs.harvard.edu/abs/1994ApJ...421..828T}
  {421, 828}

\bibitem[\protect\citeauthoryear{{Townsend} \& {Teitler}}{{Townsend} \&
  {Teitler}}{2013}]{2013Townsend}
{Townsend} R.~H.~D.,  {Teitler} S.~A.,  2013, \mn@doi [MNRAS]
  {10.1093/mnras/stt1533}, \href
  {https://ui.adsabs.harvard.edu/abs/2013MNRAS.435.3406T} {435, 3406}

\bibitem[\protect\citeauthoryear{{Turcotte}, {Richer}, {Michaud}, {Iglesias}
  \& {Rogers}}{{Turcotte} et~al.}{1998}]{1998Turcotte}
{Turcotte} S.,  {Richer} J.,  {Michaud} G.,  {Iglesias} C.~A.,   {Rogers}
  F.~J.,  1998, \mn@doi [ApJ] {10.1086/306055}, \href
  {https://ui.adsabs.harvard.edu/abs/1998ApJ...504..539T} {504, 539}

\bibitem[\protect\citeauthoryear{{Ulrich}}{{Ulrich}}{1986}]{1986Ulrich}
{Ulrich} R.~K.,  1986, \mn@doi [ApJL] {10.1086/184700}, \href
  {https://ui.adsabs.harvard.edu/abs/1986ApJ...306L..37U} {306, L37}

\bibitem[\protect\citeauthoryear{{Valle}, {Dell'Omodarme}, {Prada Moroni}  \&
  {Degl'Innocenti}}{{Valle} et~al.}{2014}]{Valle2014}
{Valle} G.,  {Dell'Omodarme} M.,  {Prada Moroni} P.~G.,   {Degl'Innocenti} S.,
  2014, \mn@doi [A \& A] {10.1051/0004-6361/201322210}, \href
  {http://adsabs.harvard.edu/abs/2014A%26A...561A.125V} {561, A125}

\bibitem[\protect\citeauthoryear{{Valle}, {Dell'Omodarme}, {Prada Moroni}  \&
  {Degl'Innocenti}}{{Valle} et~al.}{2015}]{Valle2015}
{Valle} G.,  {Dell'Omodarme} M.,  {Prada Moroni} P.~G.,   {Degl'Innocenti} S.,
  2015, \mn@doi [A \& A] {10.1051/0004-6361/201424686}, \href
  {http://adsabs.harvard.edu/abs/2015A%26A...575A..12V} {575, A12}

\bibitem[\protect\citeauthoryear{{Valle}, {Dell'Omodarme}, {Prada Moroni}  \&
  {Degl'Innocenti}}{{Valle} et~al.}{2020}]{2020Valle}
{Valle} G.,  {Dell'Omodarme} M.,  {Prada Moroni} P.~G.,   {Degl'Innocenti} S.,
  2020, \mn@doi [A\&A] {10.1051/0004-6361/201936353}, \href
  {https://ui.adsabs.harvard.edu/abs/2020A&A...635A..77V} {635, A77}

\bibitem[\protect\citeauthoryear{Verma et~al.,}{Verma
  et~al.}{2014}]{Verma_2014}
Verma K.,  et~al., 2014, \mn@doi [The Astrophysical Journal]
  {10.1088/0004-637x/790/2/138}, 790, 138

\bibitem[\protect\citeauthoryear{{Verma}, {Raodeo}, {Antia}, {Mazumdar},
  {Basu}, {Lund}  \& {Silva Aguirre}}{{Verma} et~al.}{2017}]{20Verma}
{Verma} K.,  {Raodeo} K.,  {Antia} H.~M.,  {Mazumdar} A.,  {Basu} S.,  {Lund}
  M.~N.,   {Silva Aguirre} V.,  2017, \mn@doi [ApJ] {10.3847/1538-4357/aa5da7},
  \href {https://ui.adsabs.harvard.edu/abs/2017ApJ...837...47V} {837, 47}

\bibitem[\protect\citeauthoryear{{Verma}, {Raodeo}, {Basu}, {Silva Aguirre},
  {Mazumdar}, {Mosumgaard}, {Lund}  \& {Ranadive}}{{Verma}
  et~al.}{2019}]{2019Verma}
{Verma} K.,  {Raodeo} K.,  {Basu} S.,  {Silva Aguirre} V.,  {Mazumdar} A.,
  {Mosumgaard} J.~R.,  {Lund} M.~N.,   {Ranadive} P.,  2019, \mn@doi [MNRAS]
  {10.1093/mnras/sty3374}, \href
  {https://ui.adsabs.harvard.edu/abs/2019MNRAS.483.4678V} {483, 4678}

\bibitem[\protect\citeauthoryear{{Vincenzo}, {Miglio}, {Kobayashi}, {Mackereth}
   \& {Montalban}}{{Vincenzo} et~al.}{2019}]{Vincenzo}
{Vincenzo} F.,  {Miglio} A.,  {Kobayashi} C.,  {Mackereth} J.~T.,   {Montalban}
  J.,  2019, \mn@doi [A\&A] {10.1051/0004-6361/201935886}, \href
  {https://ui.adsabs.harvard.edu/abs/2019A&A...630A.125V} {630, A125}

\bibitem[\protect\citeauthoryear{{Vrard} et~al.,}{{Vrard}
  et~al.}{2015}]{2015Vrard}
{Vrard} M.,  et~al., 2015, \mn@doi [A\&A] {10.1051/0004-6361/201425064}, \href
  {https://ui.adsabs.harvard.edu/abs/2015A&A...579A..84V} {579, A84}

\bibitem[\protect\citeauthoryear{{Weiss} \& {Schlattl}}{{Weiss} \&
  {Schlattl}}{2008}]{Weiss2008}
{Weiss} A.,  {Schlattl} H.,  2008, \mn@doi [Ap&SS] {10.1007/s10509-007-9606-5},
  \href {http://adsabs.harvard.edu/abs/2008Ap%26SS.316...99W} {316, 99}

\bibitem[\protect\citeauthoryear{{White} et~al.,}{{White}
  et~al.}{2017}]{2017Whit}
{White} T.~R.,  et~al., 2017, \mn@doi [A\&A] {10.1051/0004-6361/201628706},
  \href {https://ui.adsabs.harvard.edu/abs/2017A&A...601A..82W} {601, A82}

\makeatother
\end{thebibliography}



\section*{Appendix A}
The inferred stellar properties from the grids A, B, and C 
are published in the online version of this paper. A description
of all fields available is given in Table~\ref{GB_parameters}. A similar table is also available in machine-readable form.
\begin{table}
\centering 
\caption{Stellar Properties for our sample derived from grids A, B, and C.}
\footnotesize
\begin{tabular}{ll}        
\hline\hline 
Item & Description \\
\hline
KIC     & Kepler Input Catalog Identifier \\
Mass    & Mass in solar units \\
$\sigma_{\rm{Mass}}$   & Mass uncertainty in solar units\\
Rad  & Radius in solar units\\
$\sigma_{\rm{Rad}}$  & Radius uncertainty in solar units\\
log g    & Surface gravity in dex \\
$\sigma_{\rm{logg}}$    & Surface gravity uncertainty in dex \\
Age     &  Age in units of Myr \\
$\sigma_{\rm{Age}}$     & Age uncertainty in units of Myr \\
Lum     & Luminosity in solar units \\
$\sigma_{\rm{Lum}}$   & Luminosity uncertainty in solar units \\
Rho & Density in g/cm$^{3}$ \\
$\sigma_{\rm{Rho}}$ & Density uncertainty in g/cm$^{3}$ \\
Xi    & Fractional initial hydrogen abundance \\
$\sigma_{\rm{Xi}}$   & Fractional initial hydrogen abundance uncertainty \\
Yi    & Fractional initial helium abundance \\
$\sigma_{\rm{Yi}}$   & Fractional initial helium abundance uncertainty \\
Xs    & Fractional surface hydrogen abundance \\
$\sigma_{\rm{Xs}}$   & Fractional surface hydrogen abundance uncertainty \\
Ys    & Fractional surface helium abundance \\
$\sigma_{\rm{Ys}}$   & Fractional surface helium abundance uncertainty \\
$\alpha_{\rm{mlt}}$    & Mixing length parameter \\
$\sigma_{\alpha_{\rm{mlt}}}$    & Mixing length parameter uncertainty \\
\hline                                   
\end{tabular}
\label{GB_parameters}
\end{table}




\label{lastpage}
\end{document}